\documentclass[12pt]{article}
\usepackage{amsmath,amssymb}
\usepackage{graphicx}

\newcommand{\ol}{\overline}
\newcommand{\mq}{m_{\tilde{q}}}
\newcommand{\mg}{m_{\tilde{g}}}
\newcommand{\mr}{m_{R}}
\newcommand{\ml}{m_{L}}
\newcommand{\mKKg}{m_{g}}
\newcommand{\mKKs}{m_5}
\newcommand{\ms}{m_5}

\newcommand{\Slash}[1]{{\ooalign{\hfil/\hfil\crcr$#1$}}} 
\newcommand{\lrpartial}{\hspace{0.1em}\raisebox{1ex}{$\leftrightarrow$}\hspace{-0.85em}\raisebox{-.6ex}{$\partial$}\hspace{0.3em}}
\begin{document}
\baselineskip=16pt
\begin{titlepage}
\begin{flushright}
{\small OU-HET 714/2011}\\
\end{flushright}
\vspace*{1.2cm}

\begin{center}

{\Large\bf 
A Simple Method of Calculating Effective Operators 

} 
\lineskip .75em
\vskip 1.5cm

\normalsize
{\large N. Haba}$^1$,
{\large K. Kaneta}$^1$,
{\large S. Matsumoto}$^2$,
{\large T. Nabeshima}$^3$

\vspace{1cm}

$^1${\it Department of Physics, 
 Osaka University, Toyonaka, Osaka 560-0043, 
 Japan} \\

$^2${\it IPMU, TODIAS, University of Tokyo, Kashiwa, 277-8583, Japan}\\

$^3${\it Department of Physics, University of Toyama, Toyama 930-8555, Japan}\\


\vspace*{10mm}

{\bf Abstract}\\[5mm]
{\parbox{13cm}{\hspace{5mm}
%

It is important to obtain effective operators 
 by integrating out high energy degrees of freedom 
 in physics. 
We suggest a general method 
 of calculating 
 accurate irrelevant operators in 
 a scattering process without use of 
 equation of motions.  
By using this method, for example, 
 we will represent a complete set of 
 dimension six operators in QCD, 
 which are 
 induced from physics beyond the standard model,  
 supersymmetry and universal extra dimension.  
We will also show an example of effective anomalous 4-Fermi interactions  
 induced from a little Higgs model. 

}}

\end{center}

\end{titlepage}

\section{Introduction}

It is important to obtain effective operators 
 by integrating out high energy degrees of freedom 
 in physics. 
In a quantum field theory, 
 we can obtain effective Lagrangian 
 by integrating out high energy momentum and 
 heavy particles. 
And their effects are introduced into 
 irrelevant operators in a low energy effective theory. 
Therefore, it is important to obtain 
 effective irrelevant operators accurately 
 for a search of new physics. 
As for high energy physics, 
 search for physics beyond 
 the standard model (SM) is one of the main subjects in 
 Large Hadron Collider (LHC). 
For both theoretical and numerical analyses to 
 search new physics, 
 we stress again that it is important to obtain 
 accurate irrelevant operators because 
 they include the hints of new physics.

There have been some works 
 of listing higher dimensional operators 
 in the field content of the SM (or also including 
 right-handed neutrinos.) 
Allowed irrelevant operators in an effective theory should be 
 determined by symmetries existed in the theory. 
Let us focus on dimension six operators 
 mainly in this paper, and 
 possible dimension six operators within the SM field content
 were  listed in Ref.\cite{NPB228}. 
Meanwhile, this set was not irreducible, and 
 Refs.\cite{NPB268,Patterns} have obtained  
 a complete set systematically. 
However, Ref.\cite{Grzadkowski} insisted 
 that 80 numbers of complete set can be 
 diminishable to 59 
 by using equation of motions (EOMs). 
Which set should we use 
 in a calculation of a scattering process?
Also, when we obtain effective operator,
 how to use it 
 for a calculation of a cross section and 
 how to use EOMs in it? 
Beside a difficulty of estimating correct symmetric factors, 
 there are some above mistakable points 
 when we use irrelevant operators.

In this paper, 
 we show a general method of calculating 
 accurate irrelevant effective operators in 
 a scattering process without use of EOMs. 
By using this method, 
 we will represent 
 dimension six operators 
 induced from 
 supersymmetry (SUSY)\cite{Nilles1984},
 universal extra dimension (UED)\cite{Appelquist2001}, 
 and little Higgs model (LH)\cite{Schmaltz2005a}. 
We will show coefficients of dimension six operators 
 in QCD by integrating out 
 sparticles (KK particles) in SUSY (UED). 
We will also show coefficients of 4-Fermi operators 
 originating from anomalous interactions in LH. 
They are  
 promising candidates of beyond the SM, 
 and our analyses will shed lights on the search of
 beyond the SM. 
Our method can apply to other quantum field theories 
 in any dimensions, so that we believe
 this technique is very useful in 
 a lot of researches in physics.


\section{Method of obtaining effective operators}

Let us show a general method of obtaining accurate 
 irrelevant operators which are useful for calculating 
 a scattering processes in an effective theory. 
The effects of beyond the SM must contain 
 in higher order irrelevant operators in general. 
Thus, 
 when we integrate out 
 new particles and high momentum (physics)
 above an energy scale of $\Lambda$, 
 the effective theory ${\cal L}_{\rm eff}$ can be expanded 
 as a power of $\Lambda^{-1}$ as, 
\begin{align}
 {\cal L}_{\rm eff}
 &={\cal L}_0+\frac{1}{\Lambda}{\cal L}_1+\frac{1}{\Lambda^2}{\cal
 L}_2+\cdots,
\end{align}
where ${\cal L}_0$ is the SM Lagrangian, and  
 ${\cal L}_1$ (${\cal L}_2$) represents 
 dimension five (six) operators. 
As for dimension five operator, 
 there is only one operator written within
 the SM field contents, 
 which induces
 Majorana neutrino masses. 
Thus, let us mainly focus on non-trivial next lowest 
 operator, i.e., 
 dimension six operators 
\begin{align}
 {\cal L}_2&=\sum_ic_i{\cal O}^{(6)}_i ,
\end{align}
 where $c_i$ is a coefficient,
 and $i$ is the index of all
 possible dimension six operators allowed
 by the SM gauge symmetry.
How can we calculate dimension six operators 
 in the effective Lagrangian by integrating out
 high energy degrees of freedom? 
One correct  
 answer is to take a path integral of the full theory as 
\begin{align}
 Z&=\int {\cal D}\phi_{SM}{\cal D}\phi_h e^{iS[\phi_{SM},\phi_h]},  
\end{align}
where $\phi_{SM}$ and $\phi_h$ represent
 the SM fields and heavy fields,
 respectively.
By integrating out $\phi_h$ as 
\begin{align}
 e^{iS_{\rm eff}[\phi_{SM}]}&=\int {\cal D}\phi_h
 e^{iS[\phi_{SM},\phi_h]},
\end{align}
we can obtain an effective action, 
\begin{align}
 S_{\rm eff}[\phi_{SM}]&=S_{SM}+S_1[\phi_{SM}]+S_2[\phi_{SM}]
+\cdots,
\end{align}
where $S_1=(1/\Lambda)\int{\cal L}_1$ and
 $S_2=(1/\Lambda^2)\int{\cal L}_2$.
Coefficients of dimension six operators have been basically 
 calculated in 
 $S_2[\phi_{SM}]$, by which 
 a S-matrix element in this effective theory
 could be estimated. 
Before showing a concrete calculation method, 
 we consider a role of EOMs when we calculate 
 irrelevant operators.  

Within the SM field content, 
 one complete set of 
 ${\cal O}^{(6)}$ is explicitly listed
 in Refs\cite{NPB268,Patterns},
 where coefficients are not determined until 
 we fix a fundamental theory existing behind the SM. 
However, 
 Ref.\cite{Grzadkowski} insisted 
 the 80 numbers of dimension six operators of the complete set
 can be 
 diminishable to 59 numbers by using EOMs. 
Which set should we use 
 in a calculation of a scattering process?
To answer this question, 
 we must know correctly whether we can use EOMs in a 
 calculation of irrelevant operators 
 or not. 
Let us overview arguments of 
 Ref.\cite{Grzadkowski} at first,
 where 
 they use the fact that 
 dimension six operators 
 are related among each others through classical EOMs. 
For example, 
 let us consider  
 quark-quark-gluon interaction in 
 dimension six operators \cite{NPB228} as     
\begin{align}
{\cal O}_{qG}
&
=(\ol{q}\gamma^\mu T^a q)(iD^\nu G_{\mu\nu})^a,
\end{align}
where 
$G^a_{\mu\nu}
=\partial_\mu G^a_\nu-\partial_\nu G^a_\mu+g_sf^{abc}G^b_\mu G^c_\nu$, 
$(D^\nu G_{\mu\nu})^a
=\partial^\nu G^a_{\mu\nu}+g_sf^{abc}G^{b\nu}G^c_{\mu\nu}$, 
and EOM is given by 
\begin{align}
(D^\nu G_{\mu\nu})^a+g_s\ol{q}\gamma_\mu T^a q
&
=0. 
\end{align}
Thus, by use of EOM, 
 this operator is rewritten by 4-Fermi operator as 
\begin{align}
{\cal O}_{qG}=-ig_s(\ol{q}\gamma^\mu T^a q)
(\ol{q}'\gamma^\mu T^a q')\equiv-ig_s{\cal O}_{4F}.
\label{221}
\end{align}
It  
 means that 
 ${\cal O}_{qG}$ is not independent from ${\cal O}_{4F}$ anymore 
 through the EOMs.\footnote{ 
There are of cause other 4-Fermi operators. 
}
This is a way to 
 obtain 
 a minimal ``complete'' set of dimension six operators in Ref.\cite{Grzadkowski}
 by reduction of redundant operators through EOMs. 
However, 
 the use of EOMs must need more careful treatment, 
 and actually, 
 we must not use EOMs in  
 a calculation of a scattering process 
 (S-matrix element). 
We will explain its validity by using 
 a process, 
 $gg\rightarrow qq$, for example. 
For 
 this process, 
 dimension six operators which include interactions of 
 quark-quark-gluon and 
 quark-quark-gluon-gluon 
 must contribute 
 as in Fig.1. 
 \begin{figure}[htbp]
 \begin{center}
\includegraphics[keepaspectratio, scale=0.5,angle=-90]{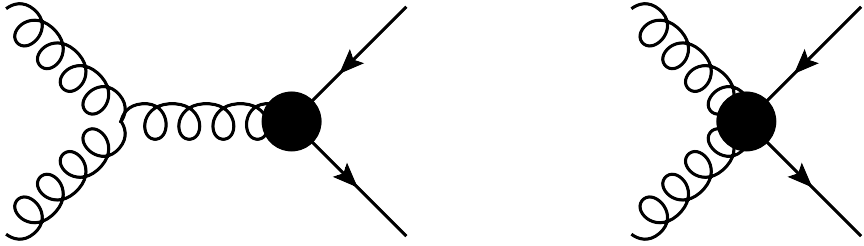}
 \end{center}
 \label{ggqq}
 \caption{
Scattering processes of $gg\rightarrow qq$: 
Black circles denote a vertex of the dimension six operator. 
There surely exist two diagrams which include 
 the irrelevant operator,  
 however, these diagram might happen to 
 vanish if we use EOMs.
}
 \end{figure} 
Among a complete set in Ref\cite{NPB268}, 
 ${\mathcal O}_{qG}$ is the only operator
 which has the interactions as  
 \begin{align}
{\cal O}_{qG}
&\supset
(qqG),\,\,(qqGG),\,\,(qqGGG). 
\end{align}
Thus, 
 a calculation of 
 $\langle qq| {\cal O}_{qG}|gg\rangle$ is seemed to be 
 enough for our goal. 
However, 
 ${\cal O}_{qG}$ is rewritten by 4-Fermi operator by use of EOMs 
 as shown above, 
 so that 
 matrix element of 
 $\langle qq| {\cal O}_{qG}|gg\rangle$ 
 might happen to vanish. 
This conclusion is quit suspicious, and 
 we must be careful to deal with EOMs 
 in a calculation of matrix element with
 general irrelevant operators
 as will be also shown 
 in Appendix A. 
We must need the accurate calculation of 
 irrelevant operators without use of EOMs. 
In this paper 
 we suggest a calculation method where we do not use 
 EOMs, 
 and calculate concrete dimension six effective
 operators by use of some new physics as the underlying theories  
 beyond the SM.

We now stand in a position of showing 
 our accurate method of calculating 
 irrelevant operators without use of 
 EOMs. 
Let us show a concrete calculation by using 
  a toy model.  
We consider a Lagrangian,  
\begin{align}
{\cal L}
&
=\ol{\psi}_l(i\Slash{\partial}-m)\psi_l+\ol{\psi}_h(i\Slash{\partial}-M)\psi_h
+\frac{1}{2}(\partial_\mu\phi)^2-\frac{1}{2}M^2\phi^2
-g\phi(\ol{\psi}_l\psi_h+\ol{\psi}_h\psi_l),
\end{align}
where 
 $\psi_l$ ($\psi_h$) denotes a light (heavy) Dirac fermion 
 and $\phi$ is a heavy real scalar with 
 $m\ll M$. 
We will obtain an effective action of $\psi_l$ after 
 integrating out heavy fields, where   
 irrelevant operators 
 must include traces of heavy particles and their interactions 
 at high energy scale. 
Let us calculate the effective action by
 integrate out $\psi_h, \phi$, 
 and show 
 dimension six operators by expanding 
 $1/M^n$. 
The effective action should be given by 
\begin{align}
e^{iS_{\rm eff}[\psi_l]}
&
=\int {\cal D}\phi{\cal D}\psi_h{\cal D}\ol{\psi}_he^{iS[\psi_l,\psi_h,\phi]},  
\end{align}
and firstly,
 by integrating out $H$, 
 it becomes 
\begin{align}
&
=\int {\cal D}\phi{\cal D}\psi_h{\cal D}\ol{\psi}_h\exp i\left\{
S_{\rm free}[\psi_l,\phi]+(\ol{\psi}_h-\ol{A}K^{-1}_0)K_0(\psi_h-K^{-1}_0A)-\ol{A}K^{-1}_0A
\right\}, \nonumber\\
&
=({\rm Det}K_0) \exp i\left\{
S_{\rm free}[\psi_l,\phi]-\ol{A}K^{-1}_0A
\right\},
\label{28}
\end{align}
where 
\begin{align}
\begin{array}{ll}
&K^{-1}_0
=\int\frac{d^4k}{(2\pi)^4}\frac{1}{\Slash{p}-M}e^{-ip(x-y)}=-i{\cal D}^{(\psi_h)}(x-y),\\
&A
=g\phi \psi_l,\,\, 
\ol{A}=g\ol{\psi}_l\phi .
\end{array}
\end{align}
The second term in Eq.(\ref{28}) is written by 
\begin{align}
-\ol{A}K^{-1}_0A
&
=-\frac{1}{2}\int d^4xd^4y\phi(x)\delta \tilde{K} (x,y)\phi(y)
\equiv-\frac{1}{2} \phi\cdot\delta\tilde{K}\cdot\phi,\\
&
\delta\tilde{K}(x,y)
\equiv-2ig^2\ol{\psi}_l(x){\cal D}^{(\psi_h)}(x-y)\psi_l(y).
\end{align}
Next step is an integration of $\phi$, which gives 
\begin{align}
e^{iS_{\rm eff}[\psi_l]}
&
=({\rm Det}K_0)\int{\cal D}\phi\exp i\left\{
S_{\rm free}[\psi_l]-\frac{1}{2}\phi\cdot(\tilde{K}_0+\delta\tilde{K})\cdot\phi\right\},\nonumber\\
&
=\left(
\frac{{\rm Det}K_0}{{\rm Det}^{\frac{1}{2}}(\tilde{K}_0+\delta\tilde{K})}
\right)
e^{iS_{\rm free}[\psi_l]},\nonumber\\
&
=\left(
\frac{{\rm Det}K_0}{{\rm Det}^{\frac{1}{2}}\tilde{K}_0}
\right)
\exp \left\{
iS_{\rm free}[\psi_l]-\frac{1}{2}{\rm Tr}\sum_{n=1}^\infty\frac{(-1)^{n+1}}{n}(\tilde{K}_0^{-1}\delta\tilde{K})^n
\right\},\\
&
\tilde{K}_0^{-1}(x,y)
\equiv\int\frac{d^4p}{(2\pi)^4}\frac{-1}{p^2-M^2}e^{-ip(x-y)}
=i{\cal D}^{(\phi)}(x-y).
\end{align}
Determinant of $K_0$ and $\tilde{K}_0$ 
 are cancelled by 
 normalization, 
 so that we finally obtain 
 the effective action of $\psi_l$ as 
\begin{align}
S_{\rm eff}[\psi_l]
&
=S_{\rm free}[\psi_l]+\frac{i}{2}{\rm
 Tr}\sum_{n=1}^\infty\frac{(-1)^{n+1}}{n}(\tilde{K}_0^{-1}\delta\tilde{K})^n.
\label{Seff}
\end{align}
Higher dimensional operators 
 are included in 
 the second term of Eq.(\ref{Seff}), 
 thus  
 dimension six operators 
 are calculated from the second order of $1/M^n$ expansion. 
A space integration of ${\cal O}(1/M^2)$
 gives 
\begin{align}
&-\frac{i}{4}\int d^4xd^4yd^4zd^4w
\tilde{K}_0^{-1}(x,y)\delta\tilde{K}(y,z)\tilde{K}_0^{-1}(z,w)
\delta\tilde{K}(w,x), \nonumber\\
&
=
-ig^4\int\frac{d^4k_1}{(2\pi)^4}\frac{d^4k_2}{(2\pi)^4}\frac{d^4k_3}{(2\pi)^4}\frac{d^4k_4}{(2\pi)^4}
(2\pi)^4\delta^4(k_1-k_2+k_3-k_4), \nonumber\\
&
\times
\frac{d^4p}{(2\pi)^4}
\frac{1}{p^2-M^2}\ol{\psi}_l(k_1)\frac{1}{\Slash{p}+\Slash{k}_1-M}\psi_l(k_2)\frac{1}{(p+k_1-k_2)^2-M^2}
\ol{\psi}_l(k_2)\frac{1}{\Slash{p}+\Slash{k}_4-M}\psi_l(k_4), 
\end{align}
and integration of all momenta $p,k_i,(i=1,\cdots,4)$  
 with 
 $k_i \ll M$ 
 induces 
 4-Fermi operators,       
\begin{align}
{\cal O}_{\rm 4F}(x)
&
=\frac{g^4}{192\pi^2}\frac{1}{M^2}
\left[
-(\ol{\psi}_l\gamma^\mu \psi_l)(\ol{\psi}_l\gamma_\mu \psi_l)+2(\ol{\psi}_l\psi_l)(\ol{\psi}_l\psi_l)
\right]. 
\end{align}
These are the dimension six operators in this model.
Notice that accurate coefficients are  
 automatically obtained without care of symmetric factors.  
Other higher order operators can be
 calculated similarly.

\section{Dimension six operators induced from new physics}

By using of the calculation method represented in the previous section, 
 we concretely calculate coefficients of 
 dimension six operators 
 induced from
 some candidates of new physics, 
 SUSY, UED, and LH. 
We obtain coefficients of dimension six operators 
 in QCD by integrating out 
 SUSY and UED particles. 
We also obtain coefficients of 4-Fermi operators 
 originating from anomalous interactions induced from LH.

\subsection{SUSY}

Let us calculate 
 coefficients of dimension six operators in QCD 
 when beyond the SM is SUSY. 
In the SUSY with $R$-parity, 
 SUSY particles propagate only inside of
 loop diagrams.  
Lagrangian of the QCD sector in SUSY SM 
 is given by   
\begin{align}
 {\cal L}=
 &{\cal L}_{SM}
 +\frac{1}{2}\ol{\tilde{g}}(i\Slash{\partial}-\mg)\tilde{g}
 -\tilde{q}^\dagger_R(\partial^2+m^2_{\tilde{q}_R})\tilde{q}_R
 -\tilde{q}^\dagger_L(\partial^2+m^2_{\tilde{q}_L})\tilde{q}_L\nonumber\\
 &+\frac{i}{2}g_sf^{abc}\ol{\tilde{g}^a}\gamma^\mu \tilde{g}^bG^c_\mu
  -ig_s\sum_q(\tilde{q}^\dagger_L \frac{\lambda}{2} ^a\lrpartial^\mu\tilde{q}_L)G^a_\mu
  -ig_s\sum_q(\tilde{q}^\dagger_R \frac{\lambda}{2} ^a\lrpartial^\mu\tilde{q}_R)G^a_\mu
 \nonumber\\
 &+g_s^2\sum_q\tilde{q}^\dagger_L \frac{\lambda}{2} ^a\frac{\lambda}{2} ^b 
  \tilde{q}_LG^a_\mu G^{b\mu}
  +g_s^2\sum_q\tilde{q}^\dagger_R \frac{\lambda}{2} ^a\frac{\lambda}{2} ^b 
  \tilde{q}_RG^a_\mu G^{b\mu}
 \nonumber\\
 &-\sqrt{2}g_s\sum_q(\tilde{q}^\dagger_L\ol{\tilde{g}^a_R}\frac{\lambda}{2} ^aq_L+{\rm h.c.})
  +\sqrt{2}g_s\sum_q(\tilde{q}^\dagger_R\ol{\tilde{g}^a_L}\frac{\lambda}{2} ^aq_R+{\rm h.c.})
 . \label{susyL}
\end{align}
where 
 $\sum_q$ represents a sum over all flavors. 
Effective action should be obtained by integrating
 out $\tilde{q}_L,\tilde{q}_R,$ 
 and $\tilde{g}$ as 
\begin{align}
 e^{iS_{\rm eff}}&=
 \int {\cal D}\tilde{g}{\cal D}\tilde{q}_L{\cal D}\ol{\tilde{q}_L}
 {\cal D}\tilde{q}_R{\cal D}\ol{\tilde{q}_R}
 \,\, e^{iS}.
\end{align}
$S_{\rm eff}$ includes all possible irrelevant operators. 
Calculating results are 
 listed  
 in Appendix C.1, where  
 coefficients of dimension six operators, 
 4-Fermi ${\cal O}_{qqqq}$, 
 quark-quark-gluon-gluon ${\cal O}_{qqGG} $,
 and 
 quark-quark-gluon ${\cal O}_{qqG}$ 
 are represented. 
We overview explicit technique to calculate them 
 in the following discussions.

\begin{figure}
 \begin{center}
\includegraphics[keepaspectratio, scale=1]{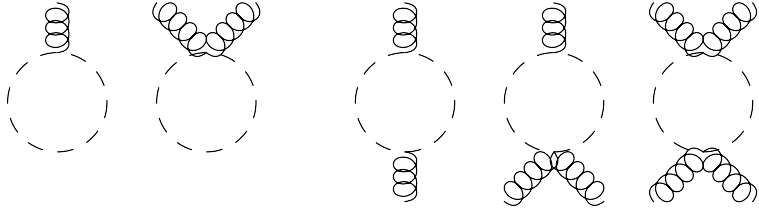} \caption{Diagrams of gluon external lines.}
 \label{gg}
 \end{center}
 \end{figure}

The first step is 
 integrating out $\tilde{q}_R$ as  
\begin{align}
 \int {\cal D}\tilde{q}_R{\cal D}\ol{\tilde{q}_R} e^{iS}&
 =\exp i\,\,
 [
 i{\rm Tr} (\log K)+B^\dagger K^{-1}B
 ], \label{qR}\\
&
\begin{array}{ll}
 K=K_0+\delta K, & \\
 K_0=(\partial^\mu\partial_\mu +m_R^2),
  &\delta K=ig_s[2G_\mu\partial^\mu+(\partial^\mu G_\mu)]-g_s^2G_\mu
  G^\mu, \\
 B=-\sqrt{2}g_s(\ol{\tilde{g}^a}\frac{\lambda}{2} ^aP_Rq),
  &B^\dagger=-\sqrt{2}g_s(\ol{q}P_L\frac{\lambda}{2} ^a\tilde{g}^a),
\end{array}
\end{align}
where ${\rm Tr} (\log K)$ includes some loop diagrams
 which have external gluon lines 
 (e.g Fig. \ref{gg}). 
The second term, $B^\dagger K^{-1}B$, 
 can be expanded by right-handed squark propagator as   
\begin{align}
B^\dagger K^{-1}B&=
B^\dagger (1+K_0^{-1}\delta K)^{-1}K_0^{-1}B\nonumber\\
&=B^\dagger K_0^{-1} B
-B^\dagger K_0^{-1} \delta K K_0^{-1} B
+B^\dagger K_0^{-1} \delta K K_0^{-1} \delta K K_0^{-1} B
+\cdots .
\end{align}
We take the expansion up to order of $g_s^4$
 in $S_{\rm eff}$.
Similarly, 
 $\tilde{q}_L$ integration can be performed, and 
 after $\tilde{q}_R, \tilde{q}_L$ integrations,
 an ``effective action'' in this stage is given by
\begin{align}
S^{(\tilde{g})}&=
\frac{1}{2}\int d^4xd^4y
\ol{\tilde{g}}(x)^a_i
\left[
\tilde{K}_0+\tilde{K}_I
\right]_{xy}^{ab}
\tilde{g}(y)^b_j, 
\end{align}
where $\tilde{K}_0$ and $\tilde{K}_I$ are
\begin{align}
\left[
\tilde{K}_0
\right]^{ab}_{xy}&\equiv
\delta^{ab}\delta^4(x-y)(i\Slash{\partial}_y-\mg)_{ij},
\end{align}
\begin{align}
\left[
\tilde{K}_I
\right]^{ab}_{xy}&\equiv
ig_sf^{abc}\delta^4(x-y)\Slash{G}^c_{ij}
-4ig_s^2\sum_{q=q_L,q_R}\ol{q}_j(y)\frac{\lambda}{2}^b
{\cal D}^{(\tilde{q})}(y-x)\frac{\lambda}{2}^aq_i(x)\nonumber\\
&
-4ig_s^3\int d^4z\sum_{q=q_L,q_R}
\ol{q}_j(y)\frac{\lambda}{2}^b{\cal D}^{(\tilde{q})}(y-z)
\{2G_\mu \partial^\mu+(\partial^\mu G_\mu)\}_z
{\cal D}^{(\tilde{q})}(z-x)\frac{\lambda}{2}^a q_i(x)\nonumber\\
&
-4ig_s^4\int d^4z d^4w\sum_{q=q_L,q_R}\ol{q}_j(y)
\ol{q}_j(y)\frac{\lambda}{2}^b {\cal D}^{(\tilde{q})}(y-z)
\{2G_\mu \partial^\mu+(\partial^\mu G_\mu)\}_z
{\cal D}^{(\tilde{q})}(z-w)\nonumber\\
&
\cdot
\{2G_\mu \partial^\mu+(\partial^\mu G_\mu)\}_w
{\cal D}^{(\tilde{q})}(w-x)
\frac{\lambda}{2}^a q_i(x)+{\cal O}(g_s^5),
\end{align}
and 
\begin{align}
{\cal D}^{(\tilde{q})}(x-y)
&
=-iK_0^{-1}
=\int\frac{d^4k}{(2\pi)^4}\frac{i}{k^2-\mq^2+i\epsilon}\mathrm{e}^{-ik\cdot
 (x-y)}. 
\end{align}
Here $i,j$ denote spinor indexes.
Next step is integrating out gluino, and 
 the final effective action is obtained as 
\begin{align}
S_{\rm eff}&=
\int d^4xd^4y\alpha(x,y)^{ab}_{ij} \tilde{K}_I(x,y)^{ab}_{ij}\nonumber\\
&
+\int d^4xd^4yd^4zd^4w
\beta(x,y,z,w)^{abcd}_{ijkl}
\tilde{K}_I(x,y)^{ab}_{ij} \tilde{K}_I(z.w)^{cd}_{kl}
+{\cal O}(\tilde{K}_I^3),\label{SUSY-Seff}
\end{align}
where 
$\alpha(x,y),\;\beta(x,y,z,w)$ consist of gluino propagator as 
\begin{align}
&\alpha(x,y)^{ab}_{ij}=
-\frac{1}{2}\delta^{ab}{\cal D}^{(\tilde{g})}(y-x)_{ji},\\
&\beta(x,y,z,w)^{abcd}_{ijkl}\nonumber\\
&
=\frac{i}{8}\delta^{ac}\delta^{bd}
\left[
C^\dagger{\cal D}^{(\tilde{g})}(x-z)
\right]_{ik}
\left[
{\cal D}^{(\tilde{g})}(w-y)C^T
\right]_{lj}
-\frac{i}{8}\delta^{ad}\delta^{bc}
{\cal D}^{(\tilde{g})}(w-x)_{li}
{\cal D}^{(\tilde{g})}(y-z)_{lj}.
\end{align}
We can know these results by differentiating
 interacting parts of the effective action as 
\begin{align}
\left.
\frac{\delta}{\delta \tilde{K}_I(x,y)^{ab}_{ij}}
e^{iS_{\rm eff}}
\right|_{\tilde{K}_I=0}
&
=i\alpha(x,y)^{ab}_{ij},\\
\left.
\frac{\delta}{\delta \tilde{K}_I(z,w)^{cd}_{kl}}
\frac{\delta}{\delta \tilde{K}_I(x,y)^{ab}_{ij}}
e^{iS_{\rm eff}}
\right|_{\tilde{K}_I=0}
&
=-\alpha(x,y)^{ab}_{ij}\alpha(z,w)^{cd}_{kl}\nonumber\\
&
+i\beta(x,y,z.w)^{abcd}_{ijkl}
+i\beta(z,w,x,y)^{cdab}_{klij}. 
\end{align}
In this stage, 
 all 
 ${\cal O}_{qqqq}$ and 
 ${\cal O}_{qqG}$
 at 1-loop level
 are included in Eq.(\ref{SUSY-Seff}) up to
 the second order of $\tilde{K}_I$.

For example, 
 the 4-Fermi operators
 all ${\cal O}_{qqqq}$ are obtained by 
 picking up 
 ${\cal O}(g_s^2)$ order terms from each 
 $\tilde{K}_I$ in  
 $\beta \tilde{K}_I\tilde{K}_I$,  
 which is given by 
 \begin{align}
&\int d^4xd^4yd^4zd^4w
\beta(x,y,z,w)^{abcd}_{ijkl}
\tilde{K}_I(x,y)^{ab}_{ij} \tilde{K}_I(z.w)^{cd}_{kl}\nonumber\\
&\supset
\int d^4xd^4yd^4zd^4w
\left[
\frac{i}{8}\delta^{ac}\delta^{bd}
\left[
C^\dagger{\cal D}^{(\tilde{g})}(x-z)
\right]_{ik}
\left[
{\cal D}^{(\tilde{g})}(w-y)C^T
\right]_{lj}\right.\nonumber\\
&\left.
-\frac{i}{8}\delta^{ad}\delta^{bc}
{\cal D}^{(\tilde{g})}(w-x)_{li}
{\cal D}^{(\tilde{g})}(y-z)_{lj}
\right]\nonumber\\
&\times
\left[
-4ig_s^2\ol{q}_j(y)\frac{\lambda}{2}^b{\cal D}^{(\tilde{q})}(y-x)\frac{\lambda}{2}^aq_i(x)
\right]
\cdot
\left[
-4ig_s^2\ol{q}'_l(w)\frac{\lambda}{2}^d{\cal D}^{(\tilde{q}')}(w-z)\frac{\lambda}{2}^cq'_k(z)
\right].\label{SUSY-4F}
\end{align}
The first term is given by 
\begin{align}
&-2ig_s^4\int d^4xd^4yd^4zd^4w\int\frac{d^4k_1}{(2\pi)^4}\frac{d^4k_2}{(2\pi)^4}\frac{d^4k_3}{(2\pi)^4}\frac{d^4k_4}{(2\pi)^4}
\frac{d^4p_1}{(2\pi)^4}\frac{d^4p_2}{(2\pi)^4}\frac{d^4p_3}{(2\pi)^4}\frac{d^4p_4}{(2\pi)^4}\nonumber\\
&\times
\left[\ol{q}_j(k_2)\frac{\lambda}{2}^b\frac{\lambda}{2}^aq_i(k_1)\right]
\cdot \left[
\ol{q}'_l(k_4)\frac{\lambda}{2}^d\frac{\lambda}{2}^cq'_k(k_3)
\right]
\frac{i[C^\dagger(\Slash{p_1}+\mg)]_{ik}}{p^2_1-\mg^2}
\frac{i[(\Slash{p_2}+\mg)C^T]_{lj}}{p^2_2-\mg^2}
\frac{i}{p^2_3-\mq^2}\frac{i}{p^2_4-m_{\tilde{q}'}^2}\nonumber\\
&\times
e^{ik_2y}e^{-ik_1x}e^{ik_4x}e^{-ik_3z}e^{-ip_1(x-z)}e^{ip_2(w-y)}e^{-ip_3(y-x)}e^{-ip_4(w-z)},
\end{align}
and by integrating out $x,y,z,w,p_2,p_3,p_4$, 
 this term becomes 
 \begin{align}
&=
-2ig_s^4\int\frac{d^4k_1}{(2\pi)^4}\frac{d^4k_2}{(2\pi)^4}\frac{d^4k_3}{(2\pi)^4}\frac{d^4k_4}{(2\pi)^4}
\delta^4(k_1-k_2+k_3-k_4)\nonumber\\
&\times
\left[\ol{q}_j(k_2)\frac{\lambda}{2}^b\frac{\lambda}{2}^aq_i(k_1)\right]
\cdot \left[
\ol{q}'_l(k_4)\frac{\lambda}{2}^d\frac{\lambda}{2}^cq'_k(k_3)
\right]
\left[
A(C^\dagger\gamma^\mu)_{ik}(\gamma^\nu C^T)_{lj}+B(C^\dagger)_{ik}(C^T)_{lj}
\right],\label{Fierz0}\\
&A=\int\frac{d^4p_1}{(2\pi)^4}\frac{p_{1\mu}(p_1+k_1-k_2)_\nu}{(p_1^2-\mg^2)[(p_1+k_1-k_2)^2-\mg^2][(p_1+k_1)^2-\mq^2][(p_1-k_1)^2-m_{\tilde{q}'}^2]},\\
&B=\int\frac{d^4p_1}{(2\pi)^4}\frac{\mg^2}{(p_1^2-\mg^2)[(p_1+k_1-k_2)^2-\mg^2][(p_1+k_1)^2-\mq^2][(p_1-k_1)^2-m_{\tilde{q}'}^2]}.
\end{align}
Here 
 $A, B$ are Feynman parameter integral, and 
 they become 
\begin{align}
A
&
\rightarrow -i\frac{6}{192\pi^2}f_1(\mq,m_{\tilde{q}}), \;\; 
B
\rightarrow i\frac{12}{192\pi^2}f_2(\mq,m_{\tilde{q}}), 
\end{align}
 when $k_i, (i=1,\cdots, 4)$ are much smaller than 
 masses of squarks and gluino. 
$f_1,f_2$ are shown in Appendix C.1, 
 and  
 the spinor can be rearranged by Fierz transformation
 of Eqs.(\ref{Fierz1}) and (\ref{Fierz2}) in Appendix B.
Necessary Fierz transformations and color factors
 are shown in Appendix B.

We can summarize all 4-Fermi operators  
 as separating 
 color singlet ${\cal O}^{(1)}_{qqqq}$ or 
 color octet ${\cal O}^{(8)}_{qqqq}$, 
 which is shown in Appendix C.1.1.  
When their chiralities are  
 $(LL)(LL)$ or $(RR)(RR)$ 
 ($L$: left-handed, $R$: right-handed), 
 the 4-Fermi operators are given by 
\begin{align}
{\cal O}^{(1)}_{qqqq}
&
= \frac{12}{192\pi^2}g_s^4
\left[
\frac{2}{9}(f_1+f_2)
\right](\ol{q}\gamma^\mu q)(\ol{q}'\gamma^\mu q'),\\
{\cal O}^{(8)}_{qqqq}
&
=\frac{12}{192\pi^2}g_s^4
\left[
-\frac{1}{3}f_1-\frac{7}{6}f_2
\right]
(\ol{q}\gamma^\mu \frac{\lambda}{2}^aq)
(\ol{q}'\gamma^\mu \frac{\lambda}{2}^aq') .
\end{align}
On the other hand, 
 when their chiralities are  
 $(LL)(RR)$ or $(RR)(LL)$, 
 the 4-Fermi operators are given by 
\begin{align}
{\cal O}^{(1)}_{qqqq}
&
= \frac{12}{192\pi^2}g_s^4
\left[
\frac{2}{9}(-f_1+f_2)
\right](\ol{q}\gamma^\mu q)(\ol{q}'\gamma^\mu q'),\\
{\cal O}^{(8)}_{qqqq}
&
=\frac{12}{192\pi^2}g_s^4
\left[
-\frac{7}{6}f_1-\frac{1}{3}f_2
\right]
(\ol{q}\gamma^\mu \frac{\lambda}{2}^aq)
(\ol{q}'\gamma^\mu \frac{\lambda}{2}^aq').
\end{align}

As for  ${\cal O}_{qqG}$, 
 vertex originates from two parts, 
 $\alpha \tilde{K}_I$ and $\beta \tilde{K}_I\tilde{K}_I$,
 as 
\begin{align}
&\int d^4xd^4y \alpha(x,y)^{ab}_{ij}\tilde{K}_I(x,y)^{ab}_{ij}
\supset
\int d^4xd^4y \left[
-\frac{1}{2}\delta^{ab}{\cal D}^{(\tilde{g})}(y-x)_{ji}
\right]\nonumber\\
&\;\;\times
\left[
-4ig_s^3\int d^4z\sum_{q=q_L,q_R}
\ol{q}_j(y)\frac{\lambda}{2}^b{\cal D}^{(\tilde{q})}(y-z)
\{2G_\mu \partial^\mu+(\partial^\mu G_\mu)\}_z
{\cal D}^{(\tilde{q})}(z-x)\frac{\lambda}{2}^a q_i(x)
\right],\label{qqG-1}\\
&\int d^4xd^4yd^4zd^4w
\beta(x,y,z,w)^{abcd}_{ijkl}
\tilde{K}_I(x,y)^{ab}_{ij} \tilde{K}_I(z.w)^{cd}_{kl}\nonumber\\
&\;\;\;\;\supset
2\int d^4xd^4yd^4zd^4w
\left[
\frac{i}{8}\delta^{ac}\delta^{bd}
\left[
C^\dagger{\cal D}^{(\tilde{g})}(x-z)
\right]_{ik}
\left[
{\cal D}^{(\tilde{g})}(w-y)C^T
\right]_{lj}\right. \nonumber\\
&
\;\;\;\;\left.
-\frac{i}{8}\delta^{ad}\delta^{bc}
{\cal D}^{(\tilde{g})}(w-x)_{li}
{\cal D}^{(\tilde{g})}(y-z)_{lj}
\right]\nonumber\\
&\;\;\;\;\times
\left[
ig_sf^{abc}\delta^4(x-y)\Slash{G}^c_{ij}
\right]\cdot
\left[
-4ig_s^2\sum_{q=q_L,q_R}\ol{q}_j(y)\frac{\lambda}{2}^b
{\cal D}^{(\tilde{q})}(y-x)\frac{\lambda}{2}^aq_i(x)
\right].\label{qqG-2}
\end{align}
Equations (\ref{qqG-1}) and (\ref{qqG-2}) include 
 not only dimension four operators but also 
 all higher dimensional operators such as  
 dimension six operator. 
Higher dimensional operators have been
 obtained by expanding the full operator by $k^2\ll \Lambda^2$, 
 where 
 $k^\mu$ and $\Lambda$ denote the momentum of
 the SM particles and SUSY particles, respectively. 
Anyhow, we can obtain all ${\cal O}_{qqG}$ 
 in the similar calculations as 4-Fermi operators, 
 which is shown Appendix C.1.2.

For  
  ${\cal O}_{qqGG}$, 
 they can be also obtained in the same manner.
${\cal O}_{qqGG}$ contains in Eq.(\ref{SUSY-Seff}), 
 and there are two contributions in 
 the first order of $\tilde{K}_I$ as 
\begin{align}
&\int d^4xd^4y \alpha(x,y)^{ab}_{ij}\tilde{K}_I(x,y)^{ab}_{ij}\nonumber\\
&\;\supset
\int d^4xd^4y \alpha(x,y)^{ab}_{ij}
\cdot 4g_s^2\int d^4z\ol{q}_j(y)\frac{\lambda}{2}^ai{\cal D}^{(\tilde{q})}(y-z)[-g_s^2G_\mu G^\mu]_z
i{\cal D}^{(\tilde{q})}(z-x)\frac{\lambda}{2}^aq_i(x),
\end{align}
\begin{align}
&\int d^4xd^4y \alpha(x,y)^{ab}_{ij}\tilde{K}_I(x,y)^{ab}_{ij}\nonumber\\
&\;\;\;\;\supset
\int d^4xd^4yd^4zd^4w \alpha(x,y)^{ab}_{ij}
(-4g_s^2)\ol{q}_j(y)\frac{\lambda}{2}^ai{\cal D}^{(\tilde{q})}(y-z)
ig_s[2G^\mu \partial_\mu+(\partial_\mu G^\mu)]_z\nonumber\\
&\;\;\;\;\times
i{\cal D}^{(\tilde{q})}(z-w)ig_s[2G^\mu \partial_\mu+(\partial_\mu G^\mu)]_w
i{\cal D}^{(\tilde{q})}(w-x)\frac{\lambda}{2}^aq_i(x).
\end{align}
Similarly, 
 there is one contribution in the second order of $\tilde{K}_I$, 
 which is shown as 
\begin{align}
&\int d^4xd^4yd^4zd^4w
\beta(x,y,z,w)^{abcd}_{ijkl}
\tilde{K}_I(x,y)^{ab}_{ij} \tilde{K}_I(z.w)^{cd}_{kl}\nonumber\\
&\;\;\;\;\supset
2\int d^4xd^4yd^4zd^4w\beta(x,y,z,w)^{abcd}_{ijkl}
[ig_sf^{abe}\Slash{G}^e_{ij}(x)\delta^4(x-y)]\nonumber\\
&\;\;\;\;
\times\int d^4z'4g_s^2\ol{q}_l(z)\frac{\lambda}{2}^di{\cal D}^{(\tilde{q})}(z-z')
\{
ig_s(2G^\mu\partial_\mu+(\partial_\mu G^\mu)
\}_{z'}
i{\cal D}^{(\tilde{q})}(z'-w)\frac{\lambda}{2}^cq_k(w)
\end{align}
There is one contribution in the third order in Eq.(\ref{SUSY-Seff}),
 $\gamma \tilde{K}_I\tilde{K}_I\tilde{K}_I$,
 where $\gamma$ is given by
\begin{align}
\gamma(x,y,z,w,u,v)^{abcdef}_{ijklmn}
&
=\frac{1}{8}
\left\{
\delta^{ad}\delta^{be}\delta^{cf}
{\cal D}^{(\tilde{g})}(w-x)_{li}{\cal D}^{(\tilde{g})}(y-u)_{jm}{\cal D}^{(\tilde{g})}(v-z)_{nk}\right.\nonumber\\
&
+\delta^{af}\delta^{bc}\delta^{ed}
{\cal D}^{(\tilde{g})}(v-x)_{ni}{\cal D}^{(\tilde{g})}(y-x)_{jk}{\cal D}^{(\tilde{g})}(w-u)_{lm}\nonumber\\
&
-\delta^{ac}\delta^{be}\delta^{df}
[C^\dagger{\cal D}^{(\tilde{g})}(z-x)]_{ki}{\cal D}^{(\tilde{g})}(y-u)_{jm}[{\cal D}^{(\tilde{g})}(v-w)C^T]_{nl}\nonumber\\
&
-\delta^{ac}\delta^{bf}\delta^{de}
[C^\dagger{\cal D}^{(\tilde{g})}(x-z)]_{ik}[{\cal D}^{(\tilde{g})}(v-y)C^T]_{nj}{\cal D}^{(\tilde{g})}(w-u)_{lm}\nonumber\\
&
-\delta^{ad}\delta^{bf}\delta^{ce}
{\cal D}^{(\tilde{g})}(w-x)_{li}[{\cal D}^{(\tilde{g})}(y-v)C^T]_{jn}[C^\dagger{\cal D}^{(\tilde{g})}(u-z)]_{mk}\nonumber\\
&
-\delta^{ae}\delta^{bc}\delta^{fd}
[C^\dagger{\cal D}^{(\tilde{g})}(u-x)]_{mi}{\cal D}^{(\tilde{g})}(y-z)_{jk}[{\cal D}^{(\tilde{g})}(w-v)C^T]_{ln}\nonumber\\
&
-\delta^{ae}\delta^{bd}\delta^{fc}
[C^\dagger{\cal D}^{(\tilde{g})}(x-u)]_{im}[{\cal D}^{(\tilde{g})}(y-w)C^T]_{lj}{\cal D}^{(\tilde{g})}(v-z)_{nk}\nonumber\\
&
\left.
-\delta^{af}\delta^{bd}\delta^{ec}
{\cal D}^{(\tilde{g})}(v-x)_{ni}[{\cal D}^{(\tilde{g})}(y-w)C^T]_{jl}[C^\dagger{\cal D}^{(\tilde{g})}(z-u)]_{km}
\right\},
\end{align}
and the third order is shown as 
\begin{align}
&\int d^4x\cdots d^4v\gamma(x,y,z,w,u,v)^{abcdef}_{ijklmn}
\tilde{K}_I(x,y)^{ab}_{ij}\tilde{K}_I(z,w)^{cd}_{kl}\tilde{K}_I(u,v)^{ef}_{mn}\nonumber\\
&\;\;\;\;\supset
\int d^4x\cdots d^4v\gamma(x,y,z,w,u,v)^{abcdef}_{ijklmn} 
\left[
(-4g_s^2)\ol{q}_j(y)\frac{\lambda}{2}^b
i{\cal D}^{(\tilde{q})}(y-x)\frac{\lambda}{2}^aq_i(x)
\right] \nonumber \\
&\;\;\;\;\;\times
\left[
ig_sf^{cdg}\Slash{G}^g(z)_{kl}\delta^4(z-w)
\right]
\left[
ig_sf^{efh}\Slash{G}^h(u)_{mn}\delta^4(u-v)
\right].\label{3rd}
\end{align}
Although $\gamma$ has eight terms in total, 
 they are all the same in Eq.(\ref{3rd})  
 since each term of $\gamma$ 
 corresponds to statistic factor in Feynman diagram. 
Notice again that we do not care about a statistic factor in each operator
 since it is automatically included in $\alpha, \beta, \gamma$.
We can obtain all ${\cal O}_{qqGG}$ induced from SUSY 
 which is shown Appendix C.1.3.


\subsection{UED}

Next, 
 we estimate QCD dimension six operators 
 induced from UED. 
The UED has KK-parity so that 
 KK particles can propagate only inside loop processes. 
As the SUSY case, 
 we can calculate dimension six operators 
 by integrating out KK particles. 

After dimensional reduction of the fifth dimensional
 compactified space of $S^1/Z_2$,
 UED Lagrangian in the 4-dimensional space-time
 is given by 
\begin{align}
{\cal L}&={\cal L}_{SM}+{\cal L}_{q^{(n)}}+{\cal L}_{G^{(n)}}+{\cal L}_{G^{(n)}_5},\nonumber\\
&
\begin{array}{ll}
{\cal L}_{q^{(n)}}
=&
\displaystyle{\sum_n}
\left[
\ol{q_L}^{(n)}i(\Slash{\partial}+ig_s\Slash{G}-m^{(n)}_L)q_L^{(n)}
+\ol{q_R}^{(n)}i(\Slash{\partial}+ig_s\Slash{G}+m^{(n)}_R)q_R^{(n)}
\right.\nonumber\\
&\left.-g_s(\ol{q}\Slash{G}^{(n)}P_Lq^{(n)}+\ol{q}i\gamma^5G^{(n)}_5P_Rq^{(n)})
-(L\longleftrightarrow R)
\right]+\cdots,\\
\\
{\cal L}_{G^{(n)}}
=&
\displaystyle{\sum_n}
\left[
-\frac{1}{4}(\partial_\mu G^{(n)a}_\nu -\partial_\nu G^{(n)a}_\mu)^2
+\frac{1}{2}m_g^{(n)2}G^{(n)a}_\mu G^{(n)a\mu}
\right.\nonumber\\
&
-\frac{1}{2}g_sf^{abc}
(\partial_\mu G^a_\nu -\partial_\nu G^a_\mu)G^{(n)b\mu}G^{(n)c\nu}
-\frac{1}{2}g_sf^{abc}(\partial_\mu G^{(n)a}_\nu -\partial_\nu G^{(n)a}_\mu)G^{(n)b\mu}G^{c\nu}
\nonumber\\
&
-\frac{1}{2}g_sf^{abc}\partial_\mu G^{(n)a}_\nu -\partial_\nu G^{(n)a}_\mu)G^{b\mu}G^{(n)c\nu}
-\frac{1}{4}f^{abc}f^{ade}
\left\{
2G^{(0)b}_\mu G^{(0)c}_\nu G^{(n)d\mu}G^{(n)e\mu}
\right.\nonumber\\
&\left.\left.
+(G^{(0)b}_\mu G^{(n)c}_\nu+G^{(n)b}_\mu G^{(0)c}_\nu)
(G^{(0)d\mu}G^{(n)e\mu}+G^{(n)d\mu}G^{(0)e\mu})
\right\}
\right]\cdots,\\
\\
{\cal L}_{G^{(n)}_5}
=&
\displaystyle{\sum_n}
\left[
\frac{1}{2}\partial_\mu G^{(n)a}_5\partial^\mu G^{(n)a}_5
-\frac{1}{2}m^2_5 G^{(n)a}_5 G^{(n)a}_5
\right.\nonumber\\
&+\left.g_sf^{abc}(m_g G^{(n)a}_\nu+\partial_\nu G^{(n)a}_5)G^{(n)b5}G^{c\nu}
-\frac{1}{2}g_s^2f^{abc}f^{ade}G^{(n)b}_5G^c_\nu G^{(n)d5}G^{e\nu}
\right]+\cdots,
\end{array}
\end{align}
where 
\begin{align}
m^{(n)}_L
&
=\frac{n}{R}+\delta m_L,\;\;\;
m^{(n)}_R
=\frac{n}{R}+\delta m_R,\nonumber\\
m^{(n)}_g
&
=\frac{n}{R}+\delta m_g,\;\;\;
m^{(n)}_5
=\frac{n}{R}+\delta m_5. \nonumber
\end{align}
Here we take a 'tHooft-Feynman gauge fixing, 
and 
 $m^{(n)}_L,m^{(n)}_R,m^{(n)}_g,m^{(n)}_5$
 are $SU(2)$ doublet KK quark mass, 
 $SU(2)$ singlet KK quark mass, KK gluon mass, 
 and KK scalar (fifth dimensional component of KK gluon) mass with
 each radiative correction, respectively.
At a tree level, these KK particles are degenerate in a minimal UED, 
 but there is a slight difference between $m^{(n)}_L$ and $m^{(n)}_R$
 when we consider radiative corrections. 
So here we show general effective operators 
 by using these general parameters. 
The effective operators of $S_{\rm eff}$ in UED
 can be calculated by the similar 
 technique of the 
 previous subsection, 
 and the results are shown in Appendix C.2. 
 
Here  
 we overview this calculation. 
By integrating out 
 KK quarks, KK scalars, and KK gluons,  
 the effective action becomes 
 \begin{align}
S_{\rm eff}
&
=\tilde{S}+\int d^4xd^4y\alpha(x,y)^{ab}_{\mu\nu} K_I(x,y)_{\mu\nu}^{ab}\nonumber\\
&
+\int d^4xd^4yd^4zd^4w\beta(x,y,z,w)^{abcd}_{\mu\nu\rho\sigma}
 K_I(x,y)_{\mu\nu}^{ab}K_I(z,w)_{\rho\sigma}^{cd}
 +\cdots \ ,
\label{UED-Seff}
\end{align}
where 
$\tilde{S}$ 
 does not include 
 KK gluons 
 that is given by  
\begin{align}
\tilde{S}
&
=-g_s^3\int d^4xd^4yd^4z {\cal D}^{(s)}(x-y)
\left[
\ol{q}_L(x)\frac{\lambda}{2}^a{\cal D}^{(L)}(x-z)\Slash{G}(z){\cal D}^{(L)}(z-y)\frac{\lambda}{2}^a q_L(y)
+(L\leftrightarrow R)
\right]\nonumber\\
&
+i\int d^4xd^4yd^4z_1d^4z_2d^4z_3\delta^4(x-y)\nonumber\\
&\times
\left[
g_sf^{bcf}G^{f\mu}(z_1)\partial_{z_1\mu}\delta^4(z_1-z_2)
+ig_s^2
\left\{
\ol{q}_L(z_1)\frac{\lambda}{2}^b{\cal D}^{(L)}(z_1-z_2)\frac{\lambda}{2}^cq_L(z)+(L\leftrightarrow R)
\right\}
\right]\nonumber\\
&\times
\left[
{\cal D}^{(s)}(x-z_1)\delta^{ab}
\right]\nonumber\\
&\times
\left[
g_sf^{deg}G^{g\nu}(z_3)\partial_{z_3\nu}\delta^4(z_3-y)
+ig_s^2
\left\{
\ol{q}_L(z_3)\frac{\lambda}{2}^b{\cal D}^{(L)}(z_3-y)\frac{\lambda}{2}^cq_L(y)+(L\leftrightarrow R)
\right\}
\right]\nonumber\\
&\times
\left[
{\cal D}^{(s)}(z_2-z_3)\delta^{cd}
\right]. 
\end{align}
Here ${\cal D}^{(L)}$, ${\cal D}^{(R)}$, and 
 ${\cal D}^{(s)}$ are  
 propagators of KK quarks and KK scalars as 
\begin{align}
{\cal D}^{(L)}
&
=\int\frac{d^4p}{(2\pi)^4}\frac{i}{\Slash{p}-\ml^{(n)}}e^{-ip(x-y)},\\
{\cal D}^{(R)}
&
=\int\frac{d^4p}{(2\pi)^4}\frac{i}{\Slash{p}+\mr^{(n)}}e^{-ip(x-y)},\\
\delta^{ab}{\cal D}^{(s)}
&
=\delta^{ab}
\int\frac{d^4p}{(2\pi)^4}\frac{i}{p^2-\mKKs^{(n)2}}e^{-ip(x-y)},
\end{align}
respectively. 
And $\alpha,\beta,K_I$ are given by 
 \begin{align}
&\alpha(x,y)^{ab}_{\mu\nu}
=\frac{1}{2}{\cal D}^{(g)}(x-y)\delta^{ab}g_{\mu\nu}
\equiv
\frac{1}{2}\delta^{ab}g_{\mu\nu}\int\frac{d^4p}{(2\pi)^4}\frac{-i}{p^2-\mKKg^2}e^{-i(x-y)},\\
&\beta(x,y,z,w)^{abcd}_{\mu\nu\rho\sigma}\nonumber\\
&
=\frac{i}{8}\left[
{\cal D}^{(g)}(x-z){\cal D}^{(g)}(y-w)\delta^{ac}\delta^{bd}g_{\mu\rho}g_{\nu\sigma}
+{\cal D}^{(g)}(x-w){\cal D}^{(g)}(y-z)\delta^{ad}\delta^{bc}g_{\mu\sigma}g_{\nu\rho}
\right],
\end{align}
\begin{align}
&K_I(x,y)^{ab}_{\mu\nu}
=-2g_sf^{abc}
\left[
\partial^\mu G^{c\nu}(x)+G^{c\mu}(x)\partial^\nu_x-g^{\mu\nu}G^{c\sigma}(x)\partial_{x\sigma}
\right]\delta^4(x-y)\nonumber\\
&
+2ig_s^2\left[
\ol{q}_L(x)\gamma^\mu\frac{\lambda}{2}^a{\cal D}^{(L)}(x-y)
\gamma^\nu\frac{\lambda}{2}^bq_L(y)
+(L\leftrightarrow R)
\right]\nonumber\\
&
+2g_s^3\int d^4z
\left[
\ol{q}_L(x)\gamma^\mu\frac{\lambda}{2}^a
{\cal D}^{(L)}(x-z)\Slash{G}(z){\cal D}^{(L)}(z-y)
\gamma^\nu\frac{\lambda}{2}^bq_L(y)
+(L\leftrightarrow R)
\right]\nonumber
\end{align}
\begin{align}
&
-ig_s^4\int d^4zd^4w\nonumber\\
&\times
\left[
\left(
\ol{q}_L(x)\gamma^\mu\frac{\lambda}{2}^a{\cal D}^{(L)}(x-z)\frac{\lambda}{2}^cq_L(z)
\right)
{\cal D}^{(s)}(z-w)
\left(
\ol{q}_L(y)\gamma^\nu\frac{\lambda}{2}^b{\cal D}^{(L)}(y-w)\frac{\lambda}{2}^cq_L(w)
\right)\right.\nonumber\\
&
+\left(
\ol{q}_L(z)\frac{\lambda}{2}^c{\cal D}^{(L)}(z-x)\gamma^\mu\frac{\lambda}{2}^aq_L(x)
\right)
{\cal D}^{(s)}(z-w)
\left(
\ol{q}_L(w)\frac{\lambda}{2}^c{\cal D}^{(L)}(w-y)\gamma^\nu\frac{\lambda}{2}^bq_L(y)
\right)\nonumber\\
&
-2\left(
\ol{q}_L(x)\gamma^\mu\frac{\lambda}{2}^a{\cal D}^{(L)}(x-z)\frac{\lambda}{2}^cq_L(z)
\right)
{\cal D}^{(s)}(z-w)
\left(
\ol{q}_L(w)\frac{\lambda}{2}^c{\cal D}^{(L)}(w-y)\gamma^\nu\frac{\lambda}{2}^bq_L(y)
\right)\nonumber\\
&
-\left(
\ol{q}_L(x)\gamma^\mu\frac{\lambda}{2}^a{\cal D}^{(L)}(x-z)\frac{\lambda}{2}^cq_L(z)
\right)
{\cal D}^{(s)}(z-w)
\left(
\ol{q}_R(y)\gamma^\nu\frac{\lambda}{2}^b{\cal D}^{(R)}(y-w)\frac{\lambda}{2}^cq_R(w)
\right)\nonumber\\
&
-\left(
\ol{q}_L(z)\frac{\lambda}{2}^c{\cal D}^{(L)}(z-x)\gamma^\mu\frac{\lambda}{2}^aq_L(x)
\right)
{\cal D}^{(s)}(z-w)
\left(
\ol{q}_R(w)\frac{\lambda}{2}^c{\cal D}^{(R)}(w-y)\gamma^\nu\frac{\lambda}{2}^bq_R(y)
\right)\nonumber\\
&
+\left(
\ol{q}_L(x)\gamma^\mu\frac{\lambda}{2}^a{\cal D}^{(L)}(x-z)\frac{\lambda}{2}^cq_L(z)
\right)
{\cal D}^{(s)}(z-w)
\left(
\ol{q}_R(w)\frac{\lambda}{2}^c{\cal D}^{(R)}(w-y)\gamma^\nu\frac{\lambda}{2}^bq_R(y)
\right)\nonumber\\
&
+\left(
\ol{q}_L(w)\frac{\lambda}{2}^c{\cal D}^{(L)}(w-y)\gamma^\nu\frac{\lambda}{2}^bq_L(y)
\right)
{\cal D}^{(s)}(z-w)
\left(
\ol{q}_R(x)\gamma^\mu\frac{\lambda}{2}^a{\cal D}^{(R)}(x-z)\frac{\lambda}{2}^cq_R(z)
\right)\nonumber\\
&
\left.
+(L \leftrightarrow R)
\right],
\label{UED-KI}
\end{align}
respectively. 
Here $\partial^\nu_x,\partial_{x\sigma}$ in the first line  
 of Eq.(\ref{UED-KI}) means derivatives of KK gluons. 

Dimension six operators induced from UED 
 can be also obtained 
 as taking KK particles are heavy enough than SM particles. 
We can show that 
 $(K_I)^{n}$ with $n=1,2,3$ 
 once contribute ${\cal O}_{qqqq}$, respectively, 
 and shown in Appendix C.2.1.   
Similarly, 
 $(K_I)^{n}$ with $n=1,3$ 
 once contribute ${\cal O}_{qqG}$, 
 respectively, 
 and   
 $(K_I)^{2}$ 
 twice contributes ${\cal O}_{qqG}$ 
 and shown in Appendix C.2.2.   
As for ${\cal O}_{qqGG}$, 
 $K_I$ four times, 
 $(K_I)^{2}$ six times, 
 $(K_I)^{3}$ five times, and 
 $(K_I)^{4}$ once 
 contribute, respectively, 
 and shown in Appendix C.2.3.    

\subsection{Little Higgs}

In this subsection,
 we calculate dimension six operators with anomalous coupling
 in LH model. 
The Lagrangian is given by
\begin{align}
{\cal L}_{\rm int}
&
=-g_s\ol{t}\Slash{G}t-\frac{2}{3}e\ol{t}\Slash{A}t-\frac{g}{\sqrt{2}}\cos\beta(\ol{b}\Slash{W}P_Lt+{\rm h.c.})
-\frac{g}{\cos\theta_W}\ol{t}\Slash{Z}\left(-\frac{2}{3}\sin^2\theta_W+\frac{1}{2}\cos^2\beta P_L\right)t,
\end{align}
where
 $\beta$ and $m_t$ are
 $\tan^{-1}\frac{\lambda^2_1}{\lambda^2_1+\lambda^2_2}\frac{v}{f}$ and 
 $\frac{\sqrt{\lambda_1\lambda_2}}{\lambda^2_1+\lambda^2_2}$,
 respectively.  
The 
 $f$ denotes the VEV of the Little Higgs. 
When we take $\lambda_1\simeq \lambda_2\simeq 1$,
 we can estimate $\cos\beta\simeq 1-\frac{v^2}{2f^2}$.
Only 4-Fermi operators have non-standard effects 
 originated from $W, Z$-boson exchanges 
 in the electroweak interaction. 
Then, below a energy scale of $M_W$,
 we can obtain ${\cal O}_{qqqq}$ as 
\begin{align}
{\cal O}^{\rm LH}
&=\frac{g^2}{2}\cos^2\beta\frac{1}{k^2-M_W^2}
(\ol{t}\gamma^\mu P_Lb)(\ol{b}\gamma^\mu P_Lt)\nonumber\\
&+\frac{g^2}{3}\tan^2\theta_W\cos^2\beta\frac{1}{k^2-M_Z^2}
(\ol{q}\gamma^\mu P_Lq)(\ol{t}\gamma^\mu P_Lt).
\end{align}
They are useful for estimating an evidence of 
 the LH. 

\section{Summary} 

It is important to obtain effective operators 
 by integrating out high energy degrees of freedom 
 in physics. 
In a quantum field theory, 
 we can obtain effective Lagrangian 
 by integrating out high energy momentum and 
 heavy particles. 
In this paper, 
 we have shown a general method of calculating 
 accurate irrelevant effective operators in 
 a scattering process without use of EOMs.  
By using this method, for example, 
 we have represented coefficients of
 dimension six operators 
 induced from SUSY, UED, and LH. 
We have shown coefficients of dimension six operators 
 in QCD by integrating out 
 sparticles (KK particles) in SUSY (UED). 
We have also shown coefficients of 4-Fermi operators 
 originating from anomalous interactions in LH. 
They are  
 promising candidates of beyond the SM, 
 and our analyses will shed lights on the search of
 beyond the SM.

Our method  
 can also give an effective action 
 by integrating out high momentum degrees of freedom of  
 massless particles. 
In this case, 
 an effective Lagrangian have 
 non-local interactions in general.  
Anyhow, our method can apply to other quantum field theories 
 in any dimensions, so that we believe
 this technique is very useful in 
 a lot of researches in physics.


\vspace{1cm}

{\large \bf Acknowledgments}\\

\noindent
This work is partially supported by Scientific Grant by Ministry of 
 Education and Science, Nos. 20540272, 20039006, and 20025004.

\appendix

\section{EOMs in a calculation of effective operators}

We must be careful to deal with EOMs 
 in 
 a calculation of a general irrelevant operator, 
 which is used for a scattering process.  
Let us consider a case that 
 an effective operator is rewritten by a form of 
\begin{align}
S_{\rm eff}[\phi]
&=\int d^4x F[\phi]\frac{\delta{\cal L}[\phi]}{\delta \phi}
\equiv F[\phi]\cdot\frac{\delta {\cal L}[\phi]}{\delta \phi},
\label{opEOM} 
\end{align}
where $F[\phi]$ is a local functional containing
 $\phi$ and $\partial \phi$. 
Notice that Eq.(\ref{opEOM}) can be also 
 obtained by a 
 transformation of $\phi$ of   
\begin{align}
\phi(x)
&\longrightarrow \phi(x)+F[\phi(x)]\label{ptTransf}.
\end{align}
The S-matrix should be unchanged under this transformation, 
 which is so-called 
 {\it equivalence theorem}\cite{Chisholm,Kamefuchi,Arzt1995}. 
Since $\frac{\delta {\cal L}[\phi]}{\delta \phi}=0$ 
 is regarded as EOM, 
 we can always eliminate operator in Eq.(\ref{opEOM}) from
 effective theory 
 by EOMs. 
In Green function's level, 
 this operation is valid whenever all fields contained in
 Eq.(\ref{opEOM}) are on-shell\cite{Polizer}. 
However, if not all fields are on-shell,
 the Green function which contain the vertex 
 Eq.(\ref{opEOM}) is not eliminated. 
We will show this situation 
 as the following discussions.

At first, we show a case that
 all fields are on-shell. 
%
The Green function relation is given by
\begin{align}
&\left< 0\left| T\left[
F[\phi]\frac{\delta {\cal L}[\phi]}{\delta \phi}(x)
\phi(x_1)\cdots\phi(x_n)
\right]\right|0\right>\nonumber\\
&
=-\sum_i\delta(x-x_i)\left<0\left|T\left[
F[\phi(x_i)]\phi(x_1)\cdots\check{\phi}(x_i)\cdots\phi(x_n)
\right]\right|0\right>,\label{SDeq}
\end{align}
where 
 the notation $\check{\phi}(x_i)$ means that 
 $\phi(x_i)$ is excluded from right-hand side. 
This equation is so-called Schwinger-Dyson equation.
We consider a scalar particle state,
 $\langle p|\phi(x)|0\rangle\neq 0$,
 and then the reduction formula gives
\begin{align}
\left< p\left|F[\phi]\frac{\delta {\cal L}[\phi]}{\delta \phi}(x)\right|p^\prime\right>
&
=\int d^4zd^4z^\prime e^{ipz}e^{-ip^\prime z^\prime}
(p^2-m^2)(p^{\prime 2}-m^2)\nonumber\\
&
\times 
\left<0\left| T\left[
\phi(z)F[\phi]\frac{\delta {\cal L}[\phi]}{\delta \phi}(x)\phi(z^\prime)
\right]\right|0\right>\nonumber\\
&
=-\int d^4zd^4z^\prime e^{ipz}e^{-ip^\prime z^\prime}
(p^2-m^2)(p^{\prime 2}-m^2)\nonumber\\
&
\times 
\big\{
\delta(x-z)\left<0\left| T\left[
F[\phi](x)\phi(z^\prime)
\right]\right|0\right>
+\delta(x-z^\prime)\left<0\left| T\left[
\phi(z)F[\phi](x)
\right]\right|0\right>
\big\}.\label{LSZ1}
\end{align}
Here we use Eq.(\ref{SDeq}) in the second equality, 
 and we can rewrite 
 the last line by using a two point function, $f_\phi(x,z)$,
 which obviously has a pole at $p^2=m^2$.
Then, Eq.(\ref{LSZ1}) can be written as
\begin{align}
{\rm (\ref{LSZ1})}&
=
\int d^4zd^4z^\prime e^{ipz}e^{-ip^\prime z^\prime}
(p^2-m^2)(p^{\prime 2}-m^2)
\big[
\delta(x-z)f_\phi(x,z^\prime)+\delta(x-z^\prime)f_\phi(z,x)
\big].
\end{align}
This means the matrix element should vanish at the on-shell.

Next,
 we show a case that
 all fields are not on-shell. 
%
The effective vertex becomes  
\begin{align}
F[\phi]\frac{\delta {\cal L}[\phi]}{\delta \phi}(x)
&
\longrightarrow
\ol{\psi}\Gamma^\mu\psi\frac{\delta {\cal L}}{\delta A^\mu}(x),
\end{align}
and we consider a process,
 $\psi_1\ol{\psi}_2\rightarrow\psi_3\ol{\psi}_4$. 
Then,
 the S-matrix element,
 \begin{align}
 &\left<\psi_3\ol{\psi}_4\left|T\left[\ol{\psi}\Gamma^\mu\psi\frac{\delta {\cal L}}{\delta A^\mu}(x) \ol{\psi}\gamma^\nu\psi A_\nu(y)\right]\right|\psi_1\ol{\psi}_2\right>,\nonumber
\end{align}
is calculated as 
\begin{align}
&\left< p_3p_4\left|T\left[\ol{\psi}\Gamma^\mu\psi\frac{\delta {\cal L}}{\delta A^\mu}(x) \,\,\ol{\psi}\gamma^\nu\psi A_\nu(y) \right]\right|p_1p_2\right>\nonumber\\
&=\int d^4z_1d^4z_2d^4z_3d^4z_4
\ol{V}(p_3;x_3)\ol{U}(p_4;x_4)(\Slash{p}_3-m_3)(\Slash{p}_4-m_4)\nonumber\\
&\times\big[
\delta(x-y)f_\psi(z_3,x)f_\psi(z_4,x)f_\psi(z_1,y)f_\psi(z_2,y)
\big]
(\Slash{p}_1-m_1)(\Slash{p}_2-m_2)U(p_2;x_2)V(p_1;x_1).
\end{align} 
Here $f_\psi(z,x)$ denotes a fermion two point function which obviously has a pole at $\Slash{p}=m$,
and $U$ and $V$ are wave functions for fermion and anti-fermion, respectively.
Thus, we can find that 
 there is a non-zero effect at
 on-shell for the process 
 containing $F[\phi]\cdot\frac{\delta {\cal L}[\phi]}{\delta \phi}$
 type effective operator. 
Consequently, when a scattering process
 has some contributions from such kind of operators, 
 we can not eliminate these effective operators by EOMs
 in the effective Lagrangian. 

Therefore, for the accurate estimations, 
 we must be careful to deal with EOMs 
 in a calculation of scattering amplitude. 
Hence, 
 in this paper, we have shown the calculation method without EOMs and 
 calculated some concrete effective 
 operators.

\section{Fierz transformation and color factor}

Here let us summarize some formulas  
 which are useful for calculations. 
Fierz transformation of 
 $\gamma$-matrix in Eq.(\ref{Fierz0}) 
 shows  
\begin{align}
(C^\dagger\gamma^\mu)_{ik}(\gamma^\nu C^T)_{lj}
&
=-\frac{1}{2}(\gamma^\mu)_{ji}(\gamma_\mu)_{lk}
-\frac{1}{2}(\gamma^\mu\gamma^5)_{ji}(\gamma_\mu\gamma^5)_{lk},\label{Fierz1}\\
(C^\dagger)_{ik}(C^T)_{lj}
&
=\frac{1}{4}(\gamma^\mu)_{ji}(\gamma_{\mu})_{lk}
-\frac{1}{4}(\gamma^\mu\gamma^5)_{ji}(\gamma_\mu\gamma^5)_{lk}\label{Fierz2}. 
\end{align}
We should notice that 
 scalar nor pseudo-scalar components do not appear by 
 the Fierz transformation, since 
 spinor components, ($i,j,k,l$), always have the same chirality 
 in each set of ($i,j$) and ($k,l$). 

The color factor becomes 
\begin{align}
\left(\frac{\lambda}{2}^b\right)_{jn}\left(\frac{\lambda}{2}^a\right)_{ni}
\left(\frac{\lambda}{2}^b\right)_{lm}\left(\frac{\lambda}{2}^a\right)_{mk}
&
=\frac{2}{9}\delta_{lk}\delta_{jk}
-\frac{1}{3}
\left(\frac{\lambda}{2}^a\right)_{lk}\left(\frac{\lambda}{2}^a\right)_{jk}\label{color1},  
\end{align}
{}from Fierz transformation of spinors.  
Next formulas are useful for  
 the second term of Eq.(\ref{SUSY-4F}), 
 whose spinor and color factor are different from 
 those of the first term.  
The  
 spinor is given by 
\begin{align}
(\gamma^\mu)_{li}(\gamma_\mu)_{jk}
&
=-\frac{1}{2}(\gamma^\mu)_{ji}(\gamma_\mu)_{lk}
+\frac{1}{2}(\gamma^\mu\gamma^5)_{ji}(\gamma_\mu\gamma^5)_{lk},\label{Fierz3}\\
\delta_{li}\delta_{jk}
&
=\frac{1}{4}(\gamma^\mu)_{ji}(\gamma_{\mu})_{lk}
+\frac{1}{4}(\gamma^\mu\gamma^5)_{ji}(\gamma_\mu\gamma^5)_{lk}\label{Fierz4}, 
\end{align}
and the color factor becomes 
\begin{align}
\left(\frac{\lambda}{2}^b\right)_{jn}\left(\frac{\lambda}{2}^a\right)_{ni}
\left(\frac{\lambda}{2}^a\right)_{lm}\left(\frac{\lambda}{2}^b\right)_{mk} 
&
=\frac{2}{9}\delta_{lk}\delta_{jk}
+\frac{7}{6}
\left(\frac{\lambda}{2}^a\right)_{lk}\left(\frac{\lambda}{2}^a\right)_{jk}\label{color2}, 
\end{align}
while $A,B$ are the same as the first term.

\section{Explicit Coefficients of dimension six operators}
\subsection{SUSY}
In SUSY SM, dimension six operators are written as follows,
\begin{align}
{\cal O}^{(1)}_{qqqq}(x)&=\frac{12g_s^4}{192\pi^2}
\left[
C_{LL}
\left(
\bar{q}\gamma^\mu P_Lq
\right)
\left(
\bar{q}'\gamma_\mu P_Lq'
\right)
+
C_{RR}
\left(
\bar{q}\gamma^\mu P_Rq
\right)
\left(
\bar{q}'\gamma_\mu P_Rq'
\right)\right.
\nonumber\\
&\qquad\qquad\left.+
C_{LR}
\left(
\bar{q}\gamma^\mu P_Lq
\right)
\left(
\bar{q}'\gamma_\mu P_Rq'
\right)
+
C_{RL}
\left(
\bar{q}\gamma^\mu P_Rq
\right)
\left(
\bar{q}'\gamma_\mu P_Lq'
\right)
\right],
\\
{\cal O}^{(8)}_{qqqq}(x)&=\frac{12g_s^4}{192\pi^2}
\left[
D_{LL}
\left(
\bar{q}T^a\gamma^\mu P_Lq
\right)
\left(
\bar{q}'T^a\gamma_\mu P_Lq'
\right)
+
D_{RR}
\left(
\bar{q}T^a\gamma^\mu P_Rq
\right)
\left(
\bar{q}'T^a\gamma_\mu P_Rq'
\right)\right.
\nonumber\\
&\qquad\qquad\left.+
D_{LR}
\left(
\bar{q}T^a\gamma^\mu P_Lq
\right)
\left(
\bar{q}'T^a\gamma_\mu P_Rq'
\right)
+
D_{RL}
\left(
\bar{q}T^a\gamma^\mu P_Rq
\right)
\left(
\bar{q}'T^a\gamma_\mu P_Lq'
\right)
\right],
\\
{\cal O}_{qqG}(x)&=\frac{g_s^3}{96\pi^2}
\int\frac{d^4k_1}{(2\pi)^4}\frac{d^4k_2}{(2\pi)^4}\frac{d^4k_3}{(2\pi)^4}
(2\pi)^4\delta^4(-k_1+k_2+k_3)
\ol{q}(k_2)T^aE^\mu_{L,R} G^a_\mu(k_3)P_{L,R}q(k_1),
\\
{\cal O}_{qqGG}(x)&=
\frac{g_s^4}{192\pi^2}
\int\frac{d^4k_1}{(2\pi)^4}\frac{d^4k_2}{(2\pi)^4}\frac{d^4k_3}{(2\pi)^4}\frac{d^4k_4}{(2\pi)^4}
(2\pi)^4\delta^4(-k_1+k_2+k_3+k_4)\nonumber\\
&\qquad\qquad\ol{q}(k_1)
\left[F^{\mu\nu}_{L,R}\delta^{ab}+H^{\mu\nu}_{L,R}T^aT^b\right]
G^a_\mu(k_2)G^b_\nu(k_3)P_{L,R}q(k_4),
\end{align}
where $E^\mu_i,\; F^{\mu\nu}_i,\; H^{\mu\nu}_i,\,(i=L,\;R)$ are
\begin{align}
E^\mu_i
&
=\{e_{1i}\Slash{k}_1+e_{2i}\Slash{k}_2\}k^\mu_1
+\{e_{1i}\Slash{k}_2+e_{2i}\Slash{k}_1\}k^\mu_2\nonumber\\
&+\{e_{3i}(k_1^2+k_2^2)-e_{4i}k_1\cdot k_2\}\gamma^\mu-e_{5i}i\epsilon^{\alpha\beta\mu\nu}\gamma_5\gamma_\nu
k_{1\alpha}k_{2\beta},\\
F^{\mu\nu}_i&=
f_{1i\alpha} i\epsilon^{\alpha\mu\nu\beta}\gamma_5\gamma_\beta
+f_{2i\alpha}g^{\mu\nu}\gamma^\alpha
+f_{3i\alpha}g^{\alpha\mu}\gamma^\nu
+f_{4i\alpha}g^{\alpha\nu}\gamma^\mu,\\
H^{\mu\nu}_i&=
h_{1i\alpha} i\epsilon^{\alpha\mu\nu\beta}\gamma_5\gamma_\beta
+h_{2i\alpha}g^{\mu\nu}\gamma^\alpha
+h_{3i\alpha}g^{\alpha\mu}\gamma^\nu
+h_{4i\alpha}g^{\alpha\nu}\gamma^\mu.
\end{align}

\subsubsection{Coefficients in ${\cal O}_{qqqq}$}
The coefficients of 4-Fermi operator are given as
 \begin{align}
C^{\rm SUSY}_{LL}&=\frac{2}{9}[f_1(m_{\tilde{q}_L},m_{\tilde{q}'_L})+f_2(m_{\tilde{q}_L},m_{\tilde{q}'_L})],\\
C^{\rm SUSY}_{RR}&=\frac{2}{9}[f_1(m_{\tilde{q}_R},m_{\tilde{q}'_R})+f_2(m_{\tilde{q}_R},m_{\tilde{q}'_R})],\\
C^{\rm SUSY}_{LR}&=-\frac{2}{9}[f_1(m_{\tilde{q}_R},m_{\tilde{q}'_L})-f_2(m_{\tilde{q}_L},m_{\tilde{q}'_R})],\\
C^{\rm SUSY}_{RL}&=-\frac{2}{9}[f_1(m_{\tilde{q}_L},m_{\tilde{q}'_R})+f_2(m_{\tilde{q}_L},m_{\tilde{q}'_R})],
\end{align}
\begin{align}
D^{\rm SUSY}_{LL}&=-\frac{1}{3}f_1(m_{\tilde{q}_L},m_{\tilde{q}'_L})-\frac{7}{6}f_2(m_{\tilde{q}_L},m_{\tilde{q}'_L}),\\
D^{\rm SUSY}_{RR}&=-\frac{1}{3}f_1(m_{\tilde{q}_R},m_{\tilde{q}'_R})-\frac{7}{6}f_2(m_{\tilde{q}_R},m_{\tilde{q}'_R}),
\end{align}
\begin{align}
D^{\rm SUSY}_{LR}&=-\frac{7}{6}f_1(m_{\tilde{q}_L},m_{\tilde{q}'_R})-\frac{1}{3}f_2(m_{\tilde{q}_L},m_{\tilde{q}'_R}),\\
D^{\rm SUSY}_{RL}&=-\frac{7}{6}f_2(m_{\tilde{q}_R},m_{\tilde{q}'_L})-\frac{1}{3}f_1(m_{\tilde{q}_R},m_{\tilde{q}'_L}),
\end{align}
\begin{align}
f_1(\mq,m_{\tilde{q}'})
&
=\int^1_0dy\int^1_0dz\frac{yz^2}{(\mg^2-\mq^2)yz+(\mq^2-m_{\tilde{q}'}^2)z+m_{\tilde{q}'}},\\
f_2(\mq,m_{\tilde{q}'})
&
=\int^1_0dy\int^1_0dz\frac{\mg^2yz^2}{[(\mg^2-\mq^2)yz+(\mq^2-m_{\tilde{q}'}^2)z+m_{\tilde{q}'}]^2}.
 \end{align}

\subsubsection{Coefficients in ${\cal O}_{qqG}$}
The coefficients of $q$-$q$-$G$ operator are given as
\begin{align}
E^\mu_L
&
\equiv E^\mu(\mq=m_{\tilde{q}_L}),\\
&
=\{e_1(m_{\tilde{q}_L})\Slash{k}_1+e_2(m_{\tilde{q}_L})\Slash{k}_2\}k^\mu_1
+\{e_1(m_{\tilde{q}_L})\Slash{k}_2+e_2(m_{\tilde{q}_L})\Slash{k}_1\}k^\mu_2\nonumber\\
&+\{e_3(m_{\tilde{q}_L})(k_1^2+k_2^2)-e_4(m_{\tilde{q}_L})k_1\cdot k_2\}\gamma^\mu
-e_5(m_{\tilde{q}_L})i\epsilon^{\alpha\beta\mu\nu}\gamma_5\gamma_\nu
k_{1\alpha}k_{2\beta},\\
E^\mu_R&=
E^\mu(m_{\tilde{q}_R}),
\end{align}
\begin{align}
&e_1(\mq)\nonumber\\
&=
\frac{107\mg^6-495\mg^4\mq^2+477\mg^2\mq^4-89\mq^6-6(\mg^6+3\mg^4\mq^2-54\mg^2\mq^4+18\mq^6)\log(\mg^2/\mq^2)}{18(\mg^2-\mq^2)^4},\\
&e_2(\mq)\nonumber\\
&=
\frac{-203\mg^6+351\mg^4\mq^2-189\mg^2\mq^4+41\mq^6+6(\mg^6+51\mg^4\mq^2-54\mg^2\mq^4+18\mq^6)\log(\mg^2/\mq^2)}{18(\mg^2-\mq^2)^4},\\
&e_3(\mq)=e_2(\mq),\\
&e_4(\mq)\nonumber\\
&=
\frac{-155\mg^6+423\mg^4\mq^2-333\mg^2\mq^4+65\mq^6+6(\mg^6+27\mg^4\mq^2-54\mg^2\mq^4+18\mq^6)\log(\mg^2/\mq^2)}{9(\mg^2-\mq^2)^4},\\
&e_5(\mq)=\frac{9(\mg^4-\mq^4-2\mg^2\mq^2\log(\mg^2/\mq^2))}{(\mg^2-\mq^2)^3}
.
\end{align}

\subsubsection{Coefficients in ${\cal O}_{qqGG}$}
The coefficients of $q$-$q$-$G$-$G$ operator are given as
\begin{align}
F^{\mu\nu}=&
f_{1\alpha} i\epsilon^{\alpha\mu\nu\beta}\gamma_5\gamma_\beta
+f_{2\alpha}g^{\mu\nu}\gamma^\alpha
+f_{3\alpha}g^{\alpha\mu}\gamma^\nu
+f_{4\alpha}g^{\alpha\nu}\gamma^\mu,\\
H^{\mu\nu}=&
h_{1\alpha} i\epsilon^{\alpha\mu\nu\beta}\gamma_5\gamma_\beta
+h_{2\alpha}g^{\mu\nu}\gamma^\alpha
+h_{3\alpha}g^{\alpha\mu}\gamma^\nu
+h_{4\alpha}g^{\alpha\nu}\gamma^\mu,\\
&f_{1\alpha}=
\left[
\frac{1}{2}S_{0\alpha}-2(P_0+Q_0)_\alpha
\right](m_a=m_{\tilde{g}},m_b=m_{\tilde{q}_i}),
\\
&f_{2\alpha}=
\left[
-3K_\alpha+\frac{1}{2}R_{3\alpha}+\frac{1}{2}S_{1\alpha}+2(P_1+Q_1)_\alpha
\right](m_a=m_{\tilde{g}},m_b=m_{\tilde{q}_i}),
\\
&f_{3\alpha}=
\left[
\frac{1}{2}R_{2\alpha}+\frac{1}{2}S_{2\alpha}+2(P_3+Q_3)_\alpha
\right](m_a=m_{\tilde{g}},m_b=m_{\tilde{q}_i}),
\\
&f_{4\alpha}=
\left[
\frac{1}{2}R_{1\alpha}+\frac{1}{2}S_{3\alpha}+2(P_2+Q_2)_\alpha
\right](m_a=m_{\tilde{g}},m_b=m_{\tilde{q}_i}),\\
&h_{1\alpha}=
\left[
12(P_0+Q_0)_\alpha
\right](m_a=m_{\tilde{g}},m_b=m_{\tilde{q}_i}),
\\
&h_{2\alpha}=
\left[
2K_\alpha-\frac{1}{3}R_{3\alpha}+12(P_1+Q_1)_\alpha
\right](m_a=m_{\tilde{g}},m_b=m_{\tilde{q}_i}),
\\
&h_{3\alpha}=
\left[
-\frac{1}{3}R_{2\alpha}+12(P_3+Q_3)_\alpha
\right](m_a=m_{\tilde{g}},m_b=m_{\tilde{q}_i}),
\\
&h_{4\alpha}=
\left[
-\frac{1}{3}R_{1\alpha}+12(P_2+Q_2)_\alpha
\right](m_a=m_{\tilde{g}},m_b=m_{\tilde{q}_i}).,
\end{align}
where $K(m_a,m_b), P(m_a,m_b), Q(m_a,m_b), R(m_a,m_b), S(m_a,m_b)$ are written in terms of a linear combination of the momentums;
\begin{align}
 K_\alpha&=a_1k_{1\alpha}+a_2k_{4\alpha},\\
P_{0\alpha}&
=(i_{11}+i_{12}+i_{13})k_{1\alpha}
+(i_{21}+i_{22}+i_{23})k_{2\alpha}
+(i_{31}+i_{32}+i_{33})k_{4\alpha},\\
P_{1\alpha}&
=(i_{11}+i_{12}-i_{13})k_{1\alpha}
+(i_{21}+i_{22}-i_{23})k_{2\alpha}
+(i_{31}+i_{32}-i_{33})k_{4\alpha},\\
P_{2\alpha}&
=(i_{11}-i_{12}+i_{13})k_{1\alpha}
+(i_{21}-i_{22}+i_{23})k_{2\alpha}
+(i_{31}-i_{32}+i_{33})k_{4\alpha},\\
P_{3\alpha}&
=(-i_{11}+i_{12}+i_{13})k_{1\alpha}
+(-i_{21}+i_{22}+i_{23})k_{2\alpha}
+(-i_{31}+i_{32}+i_{33})k_{4\alpha},\\
Q_{0\alpha}&
=(j_{11}-j_{12}+j_{13})k_{1\alpha}
+(j_{21}-j_{22}+j_{23})k_{2\alpha}
+(j_{31}-j_{32}+j_{33})k_{4\alpha},\\
Q_{1\alpha}&=Q_{0\alpha},\\
Q_{2\alpha}&
=(-j_{11}+j_{12}+j_{13})k_{1\alpha}
+(-j_{21}+j_{22}+j_{23})k_{2\alpha}
+(-j_{31}+j_{32}+j_{33})k_{4\alpha},\\
Q_{3\alpha}&
=(j_{11}+j_{12}-j_{13})k_{1\alpha}
+(j_{21}+j_{22}-j_{23})k_{2\alpha}
+(j_{31}+j_{32}-j_{33})k_{4\alpha},\\
R_{1\alpha}&
=b_{11}k_{1\alpha}+b_{21}k_{2\alpha}+b_{31}k_{4\alpha},\\
R_{2\alpha}&
=b_{12}k_{1\alpha}+b_{22}k_{2\alpha}+b_{32}k_{4\alpha},\\
R_{3\alpha}&
=b_{13}k_{1\alpha}+b_{23}k_{2\alpha}+b_{33}k_{4\alpha},
\end{align}
\begin{align}
S_{0\alpha}&
=(-f_{12}+f_{13})k_{1\alpha}
+(-f_{22}+f_{23})k_{2\alpha}
+(-f_{32}+f_{33})k_{4\alpha},\\
S_{1\alpha}&
=(f_{12}+f_{13})k_{1\alpha}
+(f_{22}+f_{23})k_{2\alpha}
+(f_{32}+f_{33})k_{4\alpha},\\
S_{2\alpha}&=S_{1\alpha},\\
S_{3\alpha}&
=(h_1-2f_{11}-f_{12}-f_{13})k_{1\alpha}
+(h_2-2f_{21}-f_{22}-f_{23})k_{2\alpha}\nonumber\\
&+(h_3-2f_{31}-f_{32}-f_{33})k_{4\alpha}.
\end{align}
These coefficients, $a_1, a_2, \cdots$, are given in terms of Feynman parameter integral as
\begin{align}
a_1(m_a,m_b)&=\int^1_0dx\int^1_0dy\frac{2 x y^2-2 y^2}{x y \left(m_a^2-m_b^2\right)+m_b^2},\\
a_2(m_a,m_b)&=\int^1_0dx\int^1_0dy\frac{2 y^2-2 y}{x y \left(m_a^2-m_b^2\right)+m_b^2},\\
i_{11}(m_a,m_b)&=-\int^1_0dx\int^1_0dy\int^1_0dz\frac{6 y^2 z^3}{z \left(m_a^2-m_b^2\right)+m_b^2},\\
i_{12}(m_a,m_b)&=i_{13}(m_a,m_b)\nonumber\\
&=\int^1_0dx\int^1_0dy\int^1_0dz\frac{6 y z^2 (1-y z)}{z \left(m_a^2-m_b^2\right)+m_b^2},\\
i_{21}(m_a,m_b)&=i_{22}(m_a,m_b)\nonumber\\
&=\int^1_0dx\int^1_0dy\int^1_0dz\frac{6 y z^2 (y z-x y z)}{z \left(m_a^2-m_b^2\right)+m_b^2},\\
i_{23}(m_a,m_b)&=\int^1_0dx\int^1_0dy\int^1_0dz\frac{6 y z^2 (-x y z+y z-1)}{z \left(m_a^2-m_b^2\right)+m_b^2},\\
i_{31}(m_a,m_b)&=\int^1_0dx\int^1_0dy\int^1_0dz\frac{6 y z^2 (y z-z+1)}{z \left(m_a^2-m_b^2\right)+m_b^2},\\
i_{32}(m_a,m_b)&=i_{33}(m_a,m_b)\nonumber\\
&=\int^1_0dx\int^1_0dy\int^1_0dz\frac{6 y z^2 (y z-z)}{z \left(m_a^2-m_b^2\right)+m_b^2},\\
j_{11}(m_a,m_b)&=j_{12}(m_a,m_b)\nonumber\\
&=\int^1_0dx\int^1_0dy\int^1_0dz\frac{6 y z^2 m_a^2 (1-y z)}{\left(z \left(m_a^2-m_b^2\right)+m_b^2\right){}^2}\\
j_{13}(m_a,m_b)&=\int^1_0dx\int^1_0dy\int^1_0dz-\frac{6 y^2 z^3 m_a^2}{\left(z \left(m_a^2-m_b^2\right)+m_b^2\right){}^2},
\\
j_{21}(m_a,m_b)&=j_{23}(m_a,m_b)\nonumber\\
&=\int^1_0dx\int^1_0dy\int^1_0dz\frac{6 y z^2 m_a^2 (y z-x y z)}{\left(z \left(m_a^2-m_b^2\right)+m_b^2\right){}^2},
\end{align}
\begin{align}
j_{22}(m_a,m_b)&=\int^1_0dx\int^1_0dy\int^1_0dz\frac{6 y z^2 m_a^2 (-x y z+y z-1)}{\left(z \left(m_a^2-m_b^2\right)+m_b^2\right){}^2},\\
j_{31}(m_a,m_b)&=j_{32}(m_a,m_b)\nonumber\\
&=\int^1_0dx\int^1_0dy\int^1_0dz\frac{6 y z^2 m_a^2 (y z-z)}{\left(z \left(m_a^2-m_b^2\right)+m_b^2\right){}^2},\\
j_{33}(m_a,m_b)&=\int^1_0dx\int^1_0dy\int^1_0dz\frac{6 y z^2 m_a^2 (y z-z+1)}{\left(z \left(m_a^2-m_b^2\right)+m_b^2\right){}^2},
\end{align}
\begin{align}
b_{11}(m_a,m_b)&=\int^1_0dx\int^1_0dy\int^1_0dz\frac{6 y z^2 (4 x y z-4 z+2)}{x y z \left(m_a^2-m_b^2\right)+m_b^2},\\
b_{12}(m_a,m_b)&=\int^1_0dx\int^1_0dy\int^1_0dz\frac{24 y z^2 (x y z-z+1)}{x y z \left(m_a^2-m_b^2\right)+m_b^2},\\
b_{13}(m_a,m_b)&=\int^1_0dx\int^1_0dy\int^1_0dz\frac{6 y z^2 (4 x y z-4 z)}{x y z \left(m_a^2-m_b^2\right)+m_b^2},\\
b_{21}(m_a,m_b)&=\int^1_0dx\int^1_0dy\int^1_0dz\frac{6 y z^2 (-4 y z+4 z-2)}{x y z \left(m_a^2-m_b^2\right)+m_b^2},\\
b_{22}(m_a,m_b)&=\int^1_0dx\int^1_0dy\int^1_0dz\frac{6 y z^2 (-4 y z+4 z-2)}{x y z \left(m_a^2-m_b^2\right)+m_b^2},\\
b_{23}(m_a,m_b)&=\int^1_0dx\int^1_0dy\int^1_0dz\frac{24 y z^2 (z-y z)}{x y z \left(m_a^2-m_b^2\right)+m_b^2},\\
b_{31}(m_a,m_b)&=\int^1_0dx\int^1_0dy\int^1_0dz\frac{24 y z^2 (z-y z)}{x y z \left(m_a^2-m_b^2\right)+m_b^2},\\
b_{32}(m_a,m_b)&=b_{33}(m_a,m_b)\nonumber\\
&=\int^1_0dx\int^1_0dy\int^1_0dz\frac{24 y (z-1) z^2}{x y z \left(m_a^2-m_b^2\right)+m_b^2},\\
f_{11}(m_a,m_b)&=\int^1_0dx\int^1_0dy\int^1_0dz\frac{6 y z^2 (2 y z-2 z+1)}{y z \left(m_a^2-m_b^2\right)+m_b^2},\\
f_{12}(m_a,m_b)&=f_{13}(m_a,m_b)\nonumber\\
&=\int^1_0dx\int^1_0dy\int^1_0dz\frac{12 y z^2 (y z-z)}{y z \left(m_a^2-m_b^2\right)+m_b^2},\\
f_{21}(m_a,m_b)&=0\\
f_{22}(m_a,m_b)&=\int^1_0dx\int^1_0dy\int^1_0dz\frac{12 y z^2 (y z-z)}{y z \left(m_a^2-m_b^2\right)+m_b^2},\\
f_{23}(m_a,m_b)&=\int^1_0dx\int^1_0dy\int^1_0dz\frac{12 y z^2 (x y z-y z+z-1)}{y z \left(m_a^2-m_b^2\right)+m_b^2},
\end{align}
\begin{align}
f_{31}(m_a,m_b)&=\int^1_0dx\int^1_0dy\int^1_0dz\frac{6 y z^2 (2 z-1)}{y z \left(m_a^2-m_b^2\right)+m_b^2},\\
f_{32}(m_a,m_b)&=f_{33}(m_a,m_b)\nonumber\\
&=\int^1_0dx\int^1_0dy\int^1_0dz\frac{12 y (z-1) z^2}{y z \left(m_a^2-m_b^2\right)+m_b^2},\\
h_{1}(m_a,m_b)&=\int^1_0dx\int^1_0dy\int^1_0dz\frac{6 y z^2 m_a^2 (2 y z-2 z+1)}{\left(y z \left(m_a^2-m_b^2\right)+m_b^2\right){}^2},\\
h_{2}(m_a,m_b)&=0,\\
h_{3}(m_a,m_b)&=\int^1_0dx\int^1_0dy\int^1_0dz\frac{6 y z^2 (2 z-1) m_a^2}{\left(y z \left(m_a^2-m_b^2\right)+m_b^2\right){}^2}.
\end{align}

\subsection{UED}
Let us list the dimension six operators of QCD in UED.
\subsubsection{Coefficients in ${\cal O}_{qqqq}$}

When $(\ol{q}q)(\ol{q}'q')$ chirality is (LL)(LL) or (RR)(RR),
4-Fermi operator is
\begin{align}
{\cal O}_{qqqq}(x)
&
=\frac{g_s^4}{192\pi^2}
\left[
f_1(\ol{q}\gamma^\mu q)(\ol{q}'\gamma_\mu q')
+f_2\left(\ol{q}\gamma^\mu \frac{\lambda}{2}^a q\right)\left(\ol{q}'\gamma^\mu \frac{\lambda}{2}^a q'\right)
\right],
\end{align}
and the coefficients $f_1, f_2$ are
\begin{align}
f_{1L}
&
=\frac{4 \left(\mKKg^4+4 \mKKg^2 \ml^2 \log
   \left(\frac{\ml}{\mKKg}\right)-\ml^4\right)}{\left(\mKKg^2-\ml^2\right)^3},\\
f_{1R}
&
=\frac{4 \left(\mKKg^4+4 \mKKg^2 \mr^2 \log
   \left(\frac{\mr}{\mKKg}\right)-\mr^4\right)}{\left(\mKKg^2-\mr^2\right)^3},
\\
f_{2L}
&
=\frac{1}{\left(\mKKg^2-\ms^2\right)
   (\ms-\ml)^2 (\ms+\ml)^2
   (\mKKg-\ml)^2 (\mKKg+\ml)^2}\nonumber\\
   &
   \left[
   18 \ml^2 \left(\mKKg^2 \left(\ms^4 \log
   \left(\frac{\ml^2}{\mKKg^2}\right)-\ml^4 \log
   \left(\frac{\mKKg^2}{\ml^2}\right)\right)+\ms^2
   \left(\mKKg^4+\ml^4\right) \log
   \left(\frac{\ms^2}{\ml^2}\right)\right.\right.\nonumber\\
   &+2 \ms^2 \mKKg^2
   \ml^2 \log
   \left(\frac{\mKKg^2}{\ms^2}\right)\nonumber\\
   &\left.\left.+(\ms^2-\mKKg^2)(\ms^2-\ml^2)(\mKKg^2-\ml^2)\right)
   +\frac{7
   \left(-\ms^4+2 \ms^2 \ml^2 \log
   \left(\frac{\ms^2}{\ml^2}\right)+\ml^4\right)}{2
   \left(\ms^2-\ml^2\right)^3}\right.\nonumber\\
   &\left.+\frac{30 \left(-\mKKg^4+2
   \mKKg^2 \ml^2 \log
   \left(\frac{\mKKg^2}{\ml^2}\right)+\ml^4\right)}{\left
   (\mKKg^2-\ml^2\right)^3}
   \right],
\end{align}
\begin{align}
   f_{2R}
&
=\frac{1}{\left(\mKKg^2-\ms^2\right)
   (\ms-\mr)^2 (\ms+\mr)^2
   (\mKKg-\mr)^2 (\mKKg+\mr)^2}\nonumber\\
   &
   \left[
   18 \mr^2 \left(\mKKg^2 \left(\ms^4 \log
   \left(\frac{\mr^2}{\mKKg^2}\right)-\mr^4 \log
   \left(\frac{\mKKg^2}{\mr^2}\right)\right)+\ms^2
   \left(\mKKg^4+\mr^4\right) \log
   \left(\frac{\ms^2}{\mr^2}\right)\right.\right.\nonumber\\
   &+2 \ms^2 \mKKg^2
   \mr^2 \log
   \left(\frac{\mKKg^2}{\ms^2}\right)\nonumber\\
   &\left.\left.+(\ms^2-\mKKg^2)(\ms^2-\mr^2)(\mKKg^2-\mr^2)\right)
   +\frac{7
   \left(-\ms^4+2 \ms^2 \mr^2 \log
   \left(\frac{\ms^2}{\mr^2}\right)+\mr^4\right)}{2
   \left(\ms^2-\mr^2\right)^3}\right.\nonumber\\
   &\left.+\frac{30 \left(-\mKKg^4+2
   \mKKg^2 \mr^2 \log
   \left(\frac{\mKKg^2}{\mr^2}\right)+\mr^4\right)}{\left
   (\mKKg^2-\mr^2\right)^3}
   \right].
\end{align}

When $(\ol{u}u)(\ol{t}t)$ chirality is (LL)(RR) or (RR)(LL),
4-Fermi operator is
\begin{align}
{\cal O}_{qqqq}(x)
&
=\frac{g_s^4}{192\pi^2}
\left[
f_3(\ol{q}\gamma^\mu q)(\ol{q}'\gamma_\mu q')
+f_4\left(\ol{q}\gamma^\mu \frac{\lambda}{2}^a q\right)\left(\ol{q}'\gamma^\mu \frac{\lambda}{2}^a q'\right)
\right],
\end{align}
and the coefficients $f_3, f_4$ are
\begin{align}
f_3(m_{q},m_{q'})
&
=-\frac{1}{\left(\mKKg^2-m_{q}^2\right)^2
   \left(\mKKg^2-m_{q'}^2\right)^2
   \left(m_{q}^2-m_{q'}^2\right)}\nonumber\\
& 
  \left[
   8 \left(2 \mKKg^2 m_{q'}^4 \left(\mKKg^2-2
   m_{q}^2\right) \log (\mKKg)-2 m_{q'}^4
   \left(\mKKg^2-m_{q}^2\right)^2 \log (m_{q'})\right.\right.\nonumber\\
&
+\mKKg^2(\mKKg^2-m_{q}^2)(\mKKg^2-m_{q'}^2)(m_{q}^2-m_{q'}^2)\nonumber\\
&
  \left.\left.   +2m_{q}^4 \left(\mKKg^4 \log
   \left(\frac{m_{q}}{\mKKg}\right)+2 \mKKg^2 m_{q'}^2
   \log \left(\frac{\mKKg}{m_{q}}\right)+m_{q'}^4 \log
   (m_{q})\right)\right)\right],\\
f_4(m_{q},m_{q'})
&
=\frac{9}{2\left(\ms^2-m_{q}^2\right)
   \left(\ms^2-m_{q'}^2\right)
   \left(m_{q}^2-m_{q'}^2\right)}\nonumber\\
   &
    \left(\frac{\left(\ms^2-m_{q}^2\right)
   \left(-\mKKg^4+4 \mKKg^4 \log (\mKKg)+m_{q'}^4-4
   m_{q'}^4 \log
   (m_{q'})\right)}{\mKKg^2-m_{q'}^2}\right.\nonumber\\
&
  -\frac{\left(\ms^2-m_{q'}^2\right) \left(-\mKKg^4+4 \mKKg^4 \log
   (\mKKg)+m_{q}^4-4 m_{q}^4 \log
   (m_{q})\right)}{\mKKg^2-m_{q}^2}\nonumber\\
&
    \left.+\frac{\left(m_{q}^2-m_{q'}^2\right) \left(-\ms^4+4 \ms^4 \log
   (\ms)+\mKKg^4-4 \mKKg^4 \log
   (\mKKg)\right)}{\ms^2-\mKKg^2}\right)\nonumber\\
&
+\frac{9}{2 \left(\ms^2-m_{q}^2\right)^2
   \left(\ms^2-m_{q'}^2\right)^2
   \left(m_{q'}^2-m_{q}^2\right)}\nonumber\\
&
    \left[
    2 \ms^2
   m_{q'}^4 \left(\ms^2-2 m_{q}^2\right) \log (\ms)-2
   m_{q'}^4 \left(\ms^2-m_{q}^2\right)^2 \log(m_{q'})\right.\nonumber
\end{align}
\begin{align}
&   +\ms^2 (\ms^2-m_{q}^2)(\ms^2-m_{q'}^2)(m_{q}^2-m_{q'}^2)\nonumber\\
&
   \left.+2 m_{q}^4 \left(\ms^4 \log
   \left(\frac{m_{q}}{\ms}\right)+2 \ms^2 m_{q'}^2
   \log \left(\frac{\ms}{m_{q}}\right)+m_{q'}^4 \log
   (m_{q})\right)\right]\nonumber\\
&
   -\frac{30}{\left(\mKKg^2-m_{q}^2\right)^2
   \left(\mKKg^2-m_{q'}^2\right)^2
   \left(m_{q}^2-m_{q'}^2\right)}\nonumber\\
&
    \left[2 \mKKg^2
   m_{q'}^4 \left(\mKKg^2-2 m_{q}^2\right) \log (\mKKg)-2
   m_{q'}^4 \left(\mKKg^2-m_{q}^2\right)^2 \log(m_{q'})\right.\nonumber\\
&   +\mKKg^2 (\mKKg^2-m_{q}^2)(\mKKg^2-m_{q'}^2)(m_{q}^2-m_{q'}^2)\nonumber\\
&
   \left.+2 m_{q}^4 \left(\mKKg^4 \log
   \left(\frac{m_{q}}{\mKKg}\right)+2 \mKKg^2 m_{q'}^2
   \log \left(\frac{\mKKg}{m_{q}}\right)+m_{q'}^4 \log
   (m_{q})\right)\right],
\end{align}
where $m_{q}, m_{q'}$ are $m_L$ or $m_R$ $(m_{q},\neq m_{q'}$).

\subsubsection{Coefficients in ${\cal O}_{qqG}$}
$q$-$q$-$G$ operator in UED takes the form as
\begin{align}
{\cal O}_{qqG}
&
=-\frac{g_s^3}{192\pi^2}\int\frac{d^4k_1}{(2\pi)^4}\frac{d^4k_2}{(2\pi)^4}\frac{d^4k_2}{(2\pi)^4}
(2\pi)^4\delta^4(-k_1+k_2+k_3)
\ol{q}(k_1)C^\mu G_\mu(k_3)q(k_2),\\
C^\mu
&
\equiv c_1i\epsilon^{\alpha\beta\mu\nu}\gamma^5\gamma_\nu+c_2g^{\mu\alpha}\gamma^\beta
+c_3 g^{\mu\beta}\gamma^\alpha+c_{4}g^{\alpha\beta}\gamma^\mu,\\
c_1
&
\equiv c_{11}k_{1\alpha}k_{2\beta},\\
c_2
&
\equiv c_{21}k_{1\alpha}k_{1\beta}+c_{22}k_{1\alpha}k_{2\beta}+c_{23}k_{2\alpha}k_{1\beta}+c_{24}k_{2\alpha}k_{2\beta},\\
c_3
&
\equiv c_{31}k_{1\alpha}k_{1\beta}+c_{32}k_{1\alpha}k_{2\beta}+c_{33}k_{2\alpha}k_{1\beta}+c_{34}k_{2\alpha}k_{2\beta},\\
c_4
&
\equiv c_{41}k_{1\alpha}k_{1\beta}+c_{42}k_{1\alpha}k_{2\beta}+c_{43}k_{2\alpha}k_{2\beta}.\\
\end{align}
Note that only $c_4$ takes the different form between the quark chirality $q_L$ and $q_R$,
because $q$-$q^{(n)}$-$G_5^{(n)}$ vertex is chiral interaction.
The coefficients are given as
\begin{align}
c_{11}
&
=6\frac{\mKKg^4-4 \mKKg^2 m_{q}^2 \log
   \left(\frac{\mKKg}{m_{q}}\right)-m_{q}^4}{\left(\mKKg^2-m_{q}^2\right)^3},\\
c_{21}
&
=6\frac{5 \mKKg^6-27 \mKKg^4 m_{q}^2+27 \mKKg^2 m_{q}^4+12 \mKKg^4
   \left(\mKKg^2-3 m_{q}^2\right) \log \left(\frac{m_{q}}{\mKKg}\right)-5
   m_{q}^6}{18 \left(\mKKg^2-m_{q}^2\right)^4},\\
c_{22}
&
=6\frac{7 \mKKg^6-27 \mKKg^2 m_{q}^4+12 \left(\mKKg^6-6 \mKKg^4 m_{q}^2+6
   \mKKg^2 m_{q}^4\right) \log \left(\frac{\mKKg}{m_{q}}\right)+20
   m_{q}^6}{18 \left(\mKKg^2-m_{q}^2\right)^4},\\
c_{23}
&
=6\frac{12 \mKKg^6 \log \left(\frac{\mKKg}{m_{q}}\right)-11 \mKKg^6+18
   \mKKg^4 m_{q}^2-9 \mKKg^2 m_{q}^4+2 m_{q}^6}{18
   \left(\mKKg^2-m_{q}^2\right)^4},
\end{align}
\begin{align}
c_{24}
&
=6\frac{12 \mKKg^6 \log \left(\frac{m_{q}}{\mKKg}\right)+11 \mKKg^6-18
   \mKKg^4 m_{q}^2+9 \mKKg^2 m_{q}^4-2 m_{q}^6}{18
   \left(\mKKg^2-m_{q}^2\right)^4},\\
c_{31}
&
=\frac{1}{3} \left(\frac{5 \mKKg^6-27 \mKKg^4 m_{q}^2+27
   \mKKg^2 m_{q}^4+12 \left(\mKKg^6-3 \mKKg^4
   m_{q}^2\right) \log \left(\frac{m_{q}}{\mKKg}\right)-5
   m_{q}^6}{\left(\mKKg^2-m_{q}^2\right)^4}\right.\nonumber\\
&
   \left.-\frac{6\left(2 \ms^6-9 \ms^4 m_{q}^2+18 \ms^2
   m_{q}^4+12 m_{q}^6 \log
   \left(\frac{m_{q}}{\ms}\right)-11
   m_{q}^6\right)}{\left(\ms^2-m_{q}^2\right)^4}\right),\\
c_{32}
&
=\frac{1}{3} \left(\frac{3 \left(7 \ms^6-36 \ms^4
   m_{q}^2+45 \ms^2 m_{q}^4+12 \left(3 \ms^2
   m_{q}^4-2 m_{q}^6\right) \log
   \left(\frac{\ms}{m_{q}}\right)-16
   m_{q}^6\right)}{\left(\ms^2-m_{q}^2\right)^4}\right.\nonumber
\end{align}
\begin{align}
&
   \left.+\frac{7 \mKKg^6-27 \mKKg^2 m_{q}^4+12 \left(\mKKg^6-6
   \mKKg^4 m_{q}^2+6 \mKKg^2 m_{q}^4\right) \log
   \left(\frac{\mKKg}{m_{q}}\right)+20
   m_{q}^6}{\left(\mKKg^2-m_{q}^2\right)^4}\right),\\
c_{33}
&
=\frac{1}{3} \left(\frac{12 \mKKg^6 \log
   \left(\frac{\mKKg}{m_{q}}\right)-11 \mKKg^6+18
   \mKKg^4 m_{q}^2-9 \mKKg^2 m_{q}^4+2
   m_{q}^6}{\left(\mKKg^2-m_{q}^2\right)^4}\right.\nonumber\\
&
   \left.-\frac{3\left(2 \ms^6-9 \ms^4 m_{q}^2+18 \ms^2
   m_{q}^4+12 m_{q}^6 \log
   \left(\frac{m_{q}}{\ms}\right)-11
   m_{q}^6\right)}{\left(\ms^2-m_{q}^2\right)^4}\right),\\
c_{34}
&
=\frac{1}{3} \left(\frac{12 \mKKg^6 \log
   \left(\frac{m_{q}}{\mKKg}\right)+11 \mKKg^6-18
   \mKKg^4 m_{q}^2+9 \mKKg^2 m_{q}^4-2
   m_{q}^6}{\left(\mKKg^2-m_{q}^2\right)^4}\right.\nonumber
   \end{align}
\begin{align}
&
   \left.-\frac{3\left(-5 \ms^6+27 \ms^4 m_{q}^2-27 \ms^2
   m_{q}^4+12 m_{q}^4 \left(m_{q}^2-3 \ms^2\right)
   \log \left(\frac{\ms}{m_{q}}\right)+5
   m_{q}^6\right)}{\left(\ms^2-m_{q}^2\right)^4}\right),\\
c_{41L}
&
=\frac{18 \mKKg \ml}{\left(\mKKg^2-\ms^2\right)^3 (\ms-\ml)
   (\ms+\ml) (\mKKg-\ml)^3
   (\mKKg+\ml)^3}\nonumber\\
&   
 \left[4 \mKKg^2 \ml^2
   \left(\ms^2-\mKKg^2\right)^3 \log (\ml)+4
   \ms^2 \mKKg^2 \log (\ms)
   \left(\mKKg^2-\ml^2\right)^3\right.\nonumber\\
&   
   +(\ms^2-\ml^2)\left((\ms^2-\mKKg^2)(\mKKg^2-\ml^2)
   \left(\ms^2 \left(\mKKg^2+\ml^2\right)-3
   \mKKg^4+\mKKg^2 \ml^2\right)\right.\nonumber\\
&   
   \left.\left.-4 \mKKg^2 \log
   (\mKKg) \left(-3 \ms^2 \mKKg^2
   \ml^2+\ms^2 \ml^2
   \left(\ms^2+\ml^2\right)+\mKKg^6\right)\right)\right]\nonumber
   \end{align}
\begin{align}
&   
   +\frac{5 \ms^6-27 \ms^4
   \ml^2+27 \ms^2 \ml^4-12 \ml^4
   \left(\ml^2-3 \ms^2\right) \log
   \left(\frac{\ms}{\ml}\right)-5 \ml^6}{2
   \left(\ms^2-\ml^2\right)^4}\nonumber\\
&
   +\frac{5 \left(-5\mKKg^6+27 \mKKg^4 \ml^2-27 \mKKg^2
   \ml^4+12 \left(\mKKg^6-3 \mKKg^4 \ml^2\right)
   \log \left(\frac{\mKKg}{\ml}\right)+5
   \ml^6\right)}{3
   \left(\mKKg^2-\ml^2\right)^4}\nonumber\\
&
   -\frac{5 \mKKg^6-27
   \mKKg^4 \ml^2+27 \mKKg^2 \ml^4+12
   \left(\mKKg^6-3 \mKKg^4 \ml^2\right) \log
   \left(\frac{\ml}{\mKKg}\right)-5 \ml^6}{3
   \left(\mKKg^2-\ml^2\right)^4},\\
c_{42L}
&
=\frac{1}{3} \left(
-\frac{108 \mKKg \ml }{\left(\mKKg^2-\ms^2\right)^3
   (\ms-\ml)^2 (\ms+\ml)^2
   (\mKKg-\ml)^2 (\mKKg+\ml)^2}\right.\nonumber\\
&
   \left[-2 \ml^4
   \left(\ms^2-\mKKg^2\right)^3 \log
   (\ml)+\left(\ml^2-\ms^2\right)
   \left((\ms^2-\mKKg^2)(\mKKg^2-\ml^2)\left(2 \ms^2
   \mKKg^2-\ml^2
   \left(\ms^2+\mKKg^2\right)\right)\right.\right.\nonumber\\
&   
   \left.\left.+2 \mKKg^2
   (\ms-\ml) (\ms+\ml) \log (\mKKg)
   \left(\ms^2 \left(\mKKg^2-2
   \ml^2\right)+\mKKg^4\right)\right)\right.\nonumber\\
&   
   \left.+2 \ms^2 \log
   (\ms) \left(\mKKg^2-\ml^2\right)^2
   \left(\ms^4+\ms^2 \mKKg^2-2 \mKKg^2
   \ml^2\right)\right]\nonumber
\end{align}
\begin{align}
&   
   +\frac{3 \left(-2\ms^6+9 \ms^4 \ml^2-18 \ms^2
   \ml^4+12 \ml^6 \log\left(\frac{\ms}{\ml}\right)+11
   \ml^6\right)}{\left(\ms^2-\ml^2\right)^4}\nonumber\\
&   
   +\frac{-12 \mKKg^6 \log \left(\frac{\mKKg}{\ml}\right)+11
   \mKKg^6-18 \mKKg^4 \ml^2+9 \mKKg^2 \ml^4-2
   \ml^6}{\left(\mKKg^2-\ml^2\right)^4}\nonumber\\
&   
   -\frac{7\mKKg^6-27 \mKKg^2 \ml^4+12 \left(\mKKg^6-6
   \mKKg^4 \ml^2+6 \mKKg^2 \ml^4\right) \log
   \left(\frac{\mKKg}{\ml}\right)+20
   \ml^6}{\left(\mKKg^2-\ml^2\right)^4}\nonumber\\
&   
   \left.+\frac{10\left(12 \mKKg^6 \log
   \left(\frac{\ml}{\mKKg}\right)+11 \mKKg^6-18
   \mKKg^4 \ml^2+9 \mKKg^2 \ml^4-2
   \ml^6\right)}{\left(\mKKg^2-\ml^2\right)^4}\right),
\end{align}

\begin{align}
c_{43L}
&
=-\frac{18 \mKKg \ml}{\left(\mKKg^2-\ms^2\right)^3
   (\ms-\ml)^3 (\ms+\ml)^3
   (\mKKg-\ml) (\mKKg+\ml)}\nonumber\\
&   
    \left(4 \ms^2 \left(\ml^2
   \left(\mKKg^2-\ms^2\right)^3 \log
   (\ml)+\mKKg^2 \left(\ms^2-\ml^2\right)^3
   \log (\mKKg)\right.\right.\nonumber\\
&   
  \left. +\log (\ms) (\ml^2-\mKKg^2)\left(\ms^6-3 \ms^2 \mKKg^2
   \ml^2+\mKKg^2 \ml^2
   \left(\mKKg^2+\ml^2\right)\right)\right)\nonumber\\
&   
   \left.+(\ms^2-\mKKg^2)(\ms^2-\ml^2)(\mKKg^2-\ml^2)
   \left(3 \ms^4-\ms^2
   \left(\mKKg^2+\ml^2\right)-\mKKg^2
   \ml^2\right)\right)\nonumber\\
&   
   +\frac{5
   \ms^6-27 \ms^4 \ml^2+27 \ms^2
   \ml^4-12 \ml^4 \left(\ml^2-3 \ms^2\right)
   \log \left(\frac{\ms}{\ml}\right)-5 \ml^6}{2
   \left(\ms^2-\ml^2\right)^4}\nonumber
\end{align}
\begin{align}
&   
   +\frac{5 \left(-5
   \mKKg^6+27 \mKKg^4 \ml^2-27 \mKKg^2
   \ml^4+12 \left(\mKKg^6-3 \mKKg^4 \ml^2\right)
   \log \left(\frac{\mKKg}{\ml}\right)+5
   \ml^6\right)}{3
   \left(\mKKg^2-\ml^2\right)^4}\nonumber\\
&   
   +\frac{-12 \mKKg^6 \log
   \left(\frac{\ml}{\mKKg}\right)-11 \mKKg^6+18
   \mKKg^4 \ml^2-9 \mKKg^2 \ml^4+2 \ml^6}{3
   \left(\mKKg^2-\ml^2\right)^4},\\
c_{41R}
&
=\frac{18 \mKKg \mr}{\left(\mKKg^2-\ms^2\right)^3 (\ms-\mr) (\ms+\mr)
   (\mKKg-\mr)^3 (\mKKg+\mr)^3}\nonumber\\
&
    \left(4 \mKKg^2 \mr^2
   \left(\mKKg^2-\ms^2\right)^3 \log (\mr)-4
   \ms^2 \mKKg^2 \log (\ms)
   \left(\mKKg^2-\mr^2\right)^3\right.\nonumber\\
&   
   +(\ms^2-\mr^2)\left(4 \mKKg^2 \log (\mKKg)
   \left(-3 \ms^2 \mKKg^2 \mr^2+\ms^2
   \mr^2
   \left(\ms^2+\mr^2\right)+\mKKg^6\right)\right.\nonumber\\
&   
   \left.\left.+(\mKKg^2-\ms^2)(\mKKg^2-\mr^2) \left(\ms^2
   \left(\mKKg^2+\mr^2\right)-3 \mKKg^4+\mKKg^2
   \mr^2\right)\right)\right)\nonumber\\
&   
   +\frac{5
   \ms^6-27 \ms^4 \mr^2+27 \ms^2
   \mr^4-12 \mr^4 \left(\mr^2-3 \ms^2\right)
   \log \left(\frac{\ms}{\mr}\right)-5 \mr^6}{2
   \left(\ms^2-\mr^2\right)^4}\nonumber\\
&   
   +\frac{5 \left(-5
   \mKKg^6+27 \mKKg^4 \mr^2-27 \mKKg^2
   \mr^4+12 \left(\mKKg^6-3 \mKKg^4 \mr^2\right)
   \log \left(\frac{\mKKg}{\mr}\right)+5
   \mr^6\right)}{3
   \left(\mKKg^2-\mr^2\right)^4}\nonumber\\
&   
   -\frac{5 \mKKg^6-27
   \mKKg^4 \mr^2+27 \mKKg^2 \mr^4+12
   \left(\mKKg^6-3 \mKKg^4 \mr^2\right) \log
   \left(\frac{\mr}{\mKKg}\right)-5 \mr^6}{3
   \left(\mKKg^2-\mr^2\right)^4},\\
c_{42R}
&
=\frac{1}{3} \left(\frac{108 \mKKg \mr}{\left(\mKKg^2-\ms^2\right)^3
   (\ms-\mr)^2 (\ms+\mr)^2
   (\mKKg-\mr)^2 (\mKKg+\mr)^2}\right.\nonumber\\
&
    \left(-2 \mr^4
   \left(\ms^2-\mKKg^2\right)^3 \log
   (\mr)\right.\nonumber\\
&   
  +\left(\mr^2-\ms^2\right)
   \left((\ms^2-\mKKg^2)(\mKKg^2-\mr^2)
    \left(2 \ms^2
   \mKKg^2-\mr^2
   \left(\ms^2+\mKKg^2\right)\right)\right.\nonumber\\
&   
   \left.+2 \mKKg^2(\ms^2-\mr^2)\log (\mKKg)
   \left(\ms^2 \left(\mKKg^2-2
   \mr^2\right)+\mKKg^4\right)\right)\nonumber\\
&   
   \left.+2 \ms^2 \log
   (\ms) \left(\mKKg^2-\mr^2\right)^2
   \left(\ms^4+\ms^2 \mKKg^2-2 \mKKg^2
   \mr^2\right)\right)\nonumber\\
&   
   +\frac{3 \left(-2\ms^6+9 \ms^4 \mr^2-18 \ms^2
   \mr^4+12 \mr^6 \log
   \left(\frac{\ms}{\mr}\right)+11
   \mr^6\right)}{\left(\ms^2-\mr^2\right)^4}\nonumber\\
&   
   +\frac{-12 \mKKg^6 \log \left(\frac{\mKKg}{\mr}\right)+11
   \mKKg^6-18 \mKKg^4 \mr^2+9 \mKKg^2 \mr^4-2
   \mr^6}{\left(\mKKg^2-\mr^2\right)^4}\nonumber\\
&
   -\frac{7
   \mKKg^6-27 \mKKg^2 \mr^4+12 \left(\mKKg^6-6
   \mKKg^4 \mr^2+6 \mKKg^2 \mr^4\right) \log
   \left(\frac{\mKKg}{\mr}\right)+20
   \mr^6}{\left(\mKKg^2-\mr^2\right)^4}\nonumber\\
&
   \left.+\frac{10
   \left(12 \mKKg^6 \log
   \left(\frac{\mr}{\mKKg}\right)+11 \mKKg^6-18
   \mKKg^4 \mr^2+9 \mKKg^2 \mr^4-2
   \mr^6\right)}{\left(\mKKg^2-\mr^2\right)^4}\right),
\end{align}
\begin{align}
c_{43R}
&
=\frac{18 \mKKg \mr}{\left(\mKKg^2-\ms^2\right)^3
   (\ms-\mr)^3 (\ms+\mr)^3
   (\mKKg-\mr) (\mKKg+\mr)}\nonumber\\
&   
    \left(4 \ms^2 \left(\mr^2
   \left(\mKKg^2-\ms^2\right)^3 \log
   (\mr)+\mKKg^2 \left(\ms^2-\mr^2\right)^3
   \log (\mKKg)\right.\right.\nonumber\\
&   
   \left.+\log (\ms) (\mr^2-\mKKg^2) \left(\ms^6-3 \ms^2 \mKKg^2
   \mr^2+\mKKg^2 \mr^2
   \left(\mKKg^2+\mr^2\right)\right)\right)\nonumber\\
&   
   \left.+(\ms^2-\mKKg^2)(\ms^2-\mr^2)(\mKKg^2-\mr^2)
   \left(3 \ms^4-\ms^2
   \left(\mKKg^2+\mr^2\right)-\mKKg^2
   \mr^2\right)\right)\nonumber\\
&   
   +\frac{5\ms^6-27 \ms^4 \mr^2+27 \ms^2
   \mr^4-12 \mr^4 \left(\mr^2-3 \ms^2\right)
   \log \left(\frac{\ms}{\mr}\right)-5 \mr^6}{2
   \left(\ms^2-\mr^2\right)^4}\nonumber\\
&   
   +\frac{5 \left(-5\mKKg^6+27 \mKKg^4 \mr^2-27 \mKKg^2
   \mr^4+12 \left(\mKKg^6-3 \mKKg^4 \mr^2\right)
   \log \left(\frac{\mKKg}{\mr}\right)+5
   \mr^6\right)}{3
   \left(\mKKg^2-\mr^2\right)^4}\nonumber\\
&   
   +\frac{-12 \mKKg^6 \log
   \left(\frac{\mr}{\mKKg}\right)-11 \mKKg^6+18
   \mKKg^4 \mr^2-9 \mKKg^2 \mr^4+2 \mr^6}{3
   \left(\mKKg^2-\mr^2\right)^4}.
\end{align}

\subsubsection{Coefficients in ${\cal O}_{qqGG}$}
The $q$-$q$-$G$-$G$ operator is written as
\begin{align}
{\cal O}_{qqGG}(x)
&=
\frac{g_s^4}{192\pi^2}
\int\frac{d^4k_1}{(2\pi)^4}\frac{d^4k_2}{(2\pi)^4}\frac{d^4k_3}{(2\pi)^4}\frac{d^4k_4}{(2\pi)^4}
(2\pi)^4\delta^4(-k_1+k_2+k_3+k_4)\nonumber\\
&
\qquad\qquad\ol{q}(k_1)
\left[F^{\mu\nu}_{L,R}\delta^{ab}+H^{\mu\nu}_{L,R}T^aT^b\right]
G^a_\mu(k_2)G^b_\nu(k_3)P_{L,R}q(k_4),
\end{align}
where $E^\mu_i,\; F^{\mu\nu}_i,\; H^{\mu\nu}_i,\,(i=L,\;R)$ are
\begin{align}
F^{\mu\nu}_i&=
f_{1i\alpha} i\epsilon^{\alpha\mu\nu\beta}\gamma_5\gamma_\beta
+f_{2i\alpha}g^{\mu\nu}\gamma^\alpha
+f_{3i\alpha}g^{\alpha\mu}\gamma^\nu
+f_{4i\alpha}g^{\alpha\nu}\gamma^\mu,\\
H^{\mu\nu}_i&=
h_{1i\alpha} i\epsilon^{\alpha\mu\nu\beta}\gamma_5\gamma_\beta
+h_{2i\alpha}g^{\mu\nu}\gamma^\alpha
+h_{3i\alpha}g^{\alpha\mu}\gamma^\nu
+h_{4i\alpha}g^{\alpha\nu}\gamma^\mu.
\end{align}
The coefficients of color singlet part, $F^{\mu\nu}_i$, are given as
\begin{align}
f_{1R\alpha}
&
=\left\{3[g_1(\mKKg,\mr)+g_4(\mKKg,\mr)+g_7(\mKKg,\mr)]+h_1(\mKKg,\mr)-h_4(\mKKg,\mr)-h_7(\mKKg,\mr)\right.
\nonumber\\
&
-\frac{1}{2}[g_1(\mKKs,\mr)+g_4(\mKKs,\mr)+g_7(\mKKs,\mr)-h_1(\mKKs,\mr)+h_4(\mKKs,\mr)-h_7(\mKKs,\mr)]\nonumber\\
&
+\frac{1}{4}[i_1(\mKKg,\mr)-2j_1(\mKKg,\mr,\mKKs)+2s_1(\mKKg,\mr)]\nonumber\\
&
\left.+\frac{1}{2}[s_8(\mKKg,\mr,\mKKs)-s_{12}(\mKKg,\mr,\mKKs)]\right\}
k_{1\alpha}
\nonumber
\end{align}
\begin{align}
&
+\left\{3[g_2(\mKKg,\mr)+g_5(\mKKg,\mr)+g_8(\mKKg,\mr)]+h_2(\mKKg,\mr)-h_5(\mKKg,\mr)-h_8(\mKKg,\mr)\right.
\nonumber\\
&
-\frac{1}{2}[g_2(\mKKs,\mr)+g_5(\mKKs,\mr)+g_8(\mKKs,\mr)-h_2(\mKKs,\mr)+h_5(\mKKs,\mr)-h_8(\mKKs,\mr)]\nonumber\\
&
+\frac{1}{4}[i_2(\mKKg,\mr)+2j_2(\mKKg,\mr,\mKKs)-2s_1(\mKKg,\mr)+n_1(\mKKg,\mr)+2n_{11}(\mKKg,\mr)]\nonumber\\
&
\left.-\frac{1}{2}[s_8(\mKKg,\mr,\mKKs)-s_{12}(\mKKg,\mr,\mKKs)]\right\}
k_{3\alpha}
\nonumber\\
&
+\left\{3[g_3(\mKKg,\mr)+g_6(\mKKg,\mr)+g_9(\mKKg,\mr)]+h_3(\mKKg,\mr)-h_6(\mKKg,\mr)-h_9(\mKKg,\mr)\right.
\nonumber\\
&
-\frac{1}{2}[g_3(\mKKs,\mr)+g_6(\mKKs,\mr)+g_9(\mKKs,\mr)-h_3(\mKKs,\mr)+h_6(\mKKs,\mr)-h_9(\mKKs,\mr)]\nonumber\\
&
+\frac{1}{4}[i_3(\mKKg,\mr)+2j_3(\mKKg,\mr,\mKKs)-2s_1(\mKKg,\mr)\nonumber\\
&
\left.-\frac{1}{2}[s_8(\mKKg,\mr,\mKKs)-s_{12}(\mKKg,\mr,\mKKs)]\right\}
k_{4\alpha},
\end{align}
\begin{align}
f_{2R\alpha}
&
=\left\{-3g_1(\mKKg,\mr)+3g_4(\mKKg,\mr)+2g_7(\mKKg,\mr)+h_1(\mKKg,\mr)-h_4(\mKKg,\mr)-h_7(\mKKg,\mr)\right.
\nonumber\\
&
-\frac{1}{2}[g_1(\mKKs,\mr)-g_4(\mKKs,\mr)+g_7(\mKKs,\mr)-h_1(\mKKs,\mr)+h_4(\mKKs,\mr)-h_7(\mKKs,\mr)]\nonumber\\
&
+\frac{1}{4}[i_4(\mKKg,\mr)-2j_1(\mKKg,\mr,\mKKs)-2j_1(\mKKs,\mr,\mKKg)]\nonumber\\
&
-3[e_1(\mKKg,\mr,\mKKs)-e_1(\mKKs,\mr,\mKKg)]\nonumber\\
&
+\frac{1}{4}[n_2(\mKKg,\mr)+2n_{12}(\mKKg,\mr)+l_1(\mr,\mKKs)-2s_2(\mKKg,\mr)]\nonumber\\
&
\left.-\frac{1}{2}[r_1(\mKKg,\mr,\mKKs)+r_{10}(\mKKg,\mr,\mKKs)-s_8(\mKKg,\mr,\mKKs)+s_{12}(\mKKg,\mr,\mKKs)]\right\}
k_{1\alpha}
\nonumber\\
&
+\left\{-3g_2 (\mKKg,\mr)+ 3 g_5(\mKKg,\mr) + 2 g_8(\mKKg,\mr)+h2 (\mKKg,\mr)- h5(\mKKg,\mr) - h_8(\mKKg,\mr)\right.\nonumber\\
&
-\frac{1}{2}[g_2(\mKKs,\mr) - g_5(\mKKs,\mr) + g_8(\mKKs,\mr) - h_2(\mKKs,\mr) + h_5(\mKKs,\mr) - h_8(\mKKs,\mr)]\nonumber\\
&
+\frac{1}{4}[i_5(\mKKs,\mr)-2j_2(\mKKg,\mr,\mKKs)-2j_2(\mKKs,\mr,\mKKg)\nonumber\\
&
+n_3(\mKKg,\mr)+2n_{13}(\mKKg,\mr)+l_2(\mr,\mKKs)-2s_3(\mKKg,\mr)]\nonumber\\
&
\left.+\frac{1}{2}[-r_2(\mKKg,\mr,\mKKs)-r_{11}(\mKKg,\mr,\mKKs)-s_8(\mKKg,\mr,\mKKs)+s_{12}(\mKKg,\mr,\mKKs)]\right\}
k_{3\alpha}
\nonumber\\
&
+\left\{-3g_3 (\mKKg,\mr)+ 3 g_6(\mKKg,\mr) + 2 g_9(\mKKg,\mr)+h3 (\mKKg,\mr)- h6(\mKKg,\mr) - h_9(\mKKg,\mr)\right.\nonumber\\
&
-\frac{1}{2}[g_3(\mKKs,\mr) - g_6(\mKKs,\mr) + g_9(\mKKs,\mr) - h_3(\mKKs,\mr) + h_6(\mKKs,\mr) - h_9(\mKKs,\mr)]\nonumber
\end{align}
\begin{align}
&
+\frac{1}{4}[i_6(\mKKs,\mr)-2j_3(\mKKg,\mr,\mKKs)-2j_3(\mKKs,\mr,\mKKg)\nonumber\\
&
-3[e_2(\mKKg,\mr,\mKKs)-e_2(\mKKs,\mr,\mKKg)]\nonumber\\
&
+\frac{1}{4}[n_4(\mKKg,\mr)+2n_{14}(\mKKg,\mr)+l_3(\mr,\mKKs)-2s_4(\mKKg,\mr)]
\nonumber\\
&
\left.+\frac{1}{2}[-r_3(\mKKg,\mr,\mKKs)-r_{12}(\mKKg,\mr,\mKKs)-s_8(\mKKg,\mr,\mKKs)+s_{12}(\mKKg,\mr,\mKKs)]\right\}
k_{4\alpha},\\
f_{3R\alpha}
&
=\left\{\frac{1}{4} [g_1(\mKKg,\mr) + 4g_4(\mKKg,\mr)-8g_7(\mKKg,\mr)+ 4 (h_1(\mKKg,\mr) + h_4(\mKKg,\mr) + h_7(\mKKg,\mr))]\right.\nonumber\\
&
-\frac{1}{2} [g_1(\mKKs,\mr) + g_4(\mKKs,\mr)- g_7(\mKKs,\mr)- h_1(\mKKs,\mr)- h_4(\mKKs,\mr) + h_7(\mKKs,\mr)]\nonumber\\
&
+ \frac{1}{4}(i_7(\mKKs,\mr) - 2 j_1(\mKKg,\mr,\mKKs)) - 3 e_1(\mKKg,\mr,\mKKs)\nonumber\\ 
&
+ \frac{1}{4} (n_5(\mKKg,\mr) + 2 n_{12}(\mKKg,\mr))+ \frac{1}{4}[l_4(\mr,\mKKs) -  2 s_2(\mKKg,\mr)]\nonumber\\
&
+ \frac{1}{2}[- t_1(\mKKg,\mr,\mKKs) - r_4(\mKKg,\mr,\mKKs) - r_{13}(\mKKg,\mr,\mKKs) \nonumber\\
&
\left.+ s_9(\mKKg,\mr,\mKKs) -  s_{13}(\mKKg,\mr,\mKKs)]\right\}
k_{1\alpha}
\nonumber\\
&
+\left\{\frac{1}{4}[g_2(\mKKg,\mr) + 4 g_5(\mKKg,\mr) - 8 g_8(\mKKg,\mr) + 4 (h_2(\mKKg,\mr) + h_5(\mKKg,\mr) + h_8(\mKKg,\mr))]\right.\nonumber\\
&
-\frac{1}{2}[g_2(\mKKs,\mr)  + g_5(\mKKs,\mr)  - g_8(\mKKs,\mr)  - h_2(\mKKs,\mr) - h_5(\mKKs,\mr) + h_8(\mKKs,\mr)]\nonumber\\
&
+ \frac{1}{4}[i_8(\mKKs,\mr) + 2 j_2(\mKKg,\mr,\mKKs) +n_6(\mKKg,\mr) + 2 n_{13}(\mKKg,\mr)]\nonumber\\
&
+\frac{1}{4}l_5(\mr,\mKKs) +\frac{1}{2}[-s_3(\mKKg,\mr) - t_2(\mKKg,\mr,\mKKs) -  r_5(\mKKg,\mr,\mKKs) - r_{14}(\mKKg,\mr,\mKKs) \nonumber\\
&
\left.+ s_{10}(\mKKg,\mr,\mKKs) -s_{14}(\mKKg,\mr,\mKKs)]\right\}
k_{3\alpha}
\nonumber\\
&
+\left\{\frac{1}{4}[g_3(\mKKg,\mr)  + 4 g_6(\mKKg,\mr)  - 8 g_9(\mKKg,\mr)  + 4 (h_3(\mKKg,\mr)  + h_6(\mKKg,\mr)  + h_9(\mKKg,\mr) )]\right.\nonumber\\
&
-\frac{1}{2}(g_3(\mKKs,\mr) + g_6(\mKKs,\mr)- g_9(\mKKs,\mr)- h_3(\mKKs,\mr) - h_6(\mKKs,\mr)+ h_9(\mKKs,\mr)) \nonumber\\
&
+ \frac{1}{4}[i_9(\mKKs,\mr) + 2 j_3(\mKKg,\mr,\mKKs)] - 3 e_2(\mKKs,\mr,\mKKg) + \frac{1}{4}[n_7(\mKKg,\mr) + 2 n_{14}(\mKKg,\mr)]\nonumber\\
&
+\frac{1}{4}l_6(\mr,\mKKs)+\frac{1}{2}[-s_4(\mKKg,\mr,\mKKs) -t_3(\mKKg,\mr,\mKKs) -r_6(\mKKg,\mr,\mKKs) - r_{15}(\mKKg,\mr,\mKKs)\nonumber\\
&
\left.+s_{11}(\mKKg,\mr,\mKKs) - s_{15}(\mKKg,\mr,\mKKs)]\right\}
k_{4\alpha},
\end{align}
\begin{align}
f_{4R\alpha}
&
=\left\{\frac{1}{4} [12 g_1(\mKKg,\mr) - 12 g_4(\mKKg,\mr) + g_7(\mKKg,\mr) \right.\nonumber\\
&+ 4 (-h_1(\mKKg,\mr) + h_4(\mKKg,\mr) + h_7(\mKKg,\mr))]\nonumber\\
&
- \frac{1}{2} (-g_1(\mKKs,\mr) + g_4(\mKKs,\mr)  + g_7(\mKKs,\mr) + h_1(\mKKs,\mr) - h_4(\mKKs,\mr) - h_7(\mKKs,\mr) ) \nonumber\\
&
+\frac{1}{4}[i_{10}(\mKKs,\mr)  - 2 j_1(\mKKg,\mr,\mKKs)] -3e_1(\mKKs,\mr,\mKKg) + \frac{1}{4}[ n_8(\mKKg,\mr) +l_7(\mr,\mKKs)]\nonumber\\
&
+\frac{1}{2}[s_5(\mKKg,\mr) + s_2(\mKKg,\mr)]+\frac{1}{2}[ -t_4(\mKKg,\mr,\mKKs) - r_7(\mKKg,\mr,\mKKs) - r_{16}(\mKKg,\mr,\mKKs)\nonumber\\
&
\left. - s_8(\mKKg,\mr,\mKKs) +s_{12}(\mKKg,\mr,\mKKs)]\right\}
k_{1\alpha}
\nonumber\\
&
+\left\{\frac{1}{4}[12 g_2(\mKKg,\mr) - 12 g_5(\mKKg,\mr) + g_8(\mKKg,\mr) \right.\nonumber\\
&+ 4 (-h_2(\mKKg,\mr) + h_5 (\mKKg,\mr)+ h_8(\mKKg,\mr))]\nonumber\\
&
 -\frac{1}{2} (-g_2(\mKKs,\mr) + g_5(\mKKs,\mr)+ g_8(\mKKs,\mr) + h_2(\mKKs,\mr)- h_5(\mKKs,\mr)- h_8(\mKKs,\mr))\nonumber\\
&
 +\frac{1}{4}[i_{11}(\mKKs,\mr) + 2 j_2(\mKKg,\mr,\mKKs) + n_9 (\mKKg,\mr)+ l_8(\mr,\mKKs)]\nonumber\\
&
+ \frac{1}{2}[s_6(\mKKg,\mr) +s_3(\mKKg,\mr)]+\frac{1}{2}[- t_5(\mKKg,\mr,\mKKs) - r_8(\mKKg,\mr,\mKKs) -  r_{17}(\mKKg,\mr,\mKKs)\nonumber\\
&
 \left.+ s_8(\mKKg,\mr,\mKKs) -s_{12}(\mKKg,\mr,\mKKs)]\right\}
k_{3\alpha}
\nonumber\\
&
+\left\{\frac{1}{4}[12 g_3(\mKKg,\mr)- 12 g_6(\mKKg,\mr) + g_9(\mKKg,\mr) \right.\nonumber\\
&+ 4 (-h_3(\mKKg,\mr) + h_6(\mKKg,\mr) + h_9(\mKKg,\mr))]\nonumber\\
&
-\frac{1}{2}(-g_3(\mKKs,\mr) + g_6(\mKKs,\mr)+ g_9(\mKKs,\mr)+ h_3(\mKKs,\mr)- h_6(\mKKs,\mr)- h_9(\mKKs,\mr))\nonumber\\
&
+\frac{1}{4}[i_{12}(\mKKs,\mr) + 2  j_3(\mKKg,\mr,\mKKs)]-3e_2(\mKKs,\mr,\mKKg)+\frac{1}{4}[n_{10}(\mKKg,\mr) + l_9(\mr,\mKKs)]\nonumber\\
&
+\frac{1}{2}[s_7(\mKKg,\mr) - s_4(\mKKg,\mr)-t_6(\mKKg,\mr)]+\frac{1}{2}[- r_9(\mKKg,\mr,\mKKs)- r_{18}(\mKKg,\mr,\mKKs)\nonumber\\
&
\left.+s_8 (\mKKg,\mr,\mKKs)- s_{12}(\mKKg,\mr,\mKKs)]\right\}
k_{4\alpha},\\
f_{1L\alpha}
&
=\left\{3[g_1(\mKKg,\ml)+g_4(\mKKg,\ml)+g_7(\mKKg,\ml)]+h_1(\mKKg,\ml)-h_4(\mKKg,\ml)-h_7(\mKKg,\ml)\right.
\nonumber\\
&
-\frac{1}{2}[g_1(\mKKs,\ml)+g_4(\mKKs,\ml)+g_7(\mKKs,\ml)-h_1(\mKKs,\ml)+h_4(\mKKs,\ml)-h_7(\mKKs,\ml)]\nonumber\\
&
+\frac{1}{4}[i_1(\mKKg,\ml)-2j_1(\mKKg,\ml,\mKKs)+2s_1(\mKKg,\ml)]\nonumber\\
&
\left.+\frac{1}{2}[-s_8(\mKKg,\ml,\mKKs)+s_{12}(\mKKg,\ml,\mKKs)]\right\}
k_{1\alpha}
\nonumber
\end{align}
\begin{align}
&
+\left\{3[g_2(\mKKg,\ml)+g_5(\mKKg,\ml)+g_8(\mKKg,\ml)]+h_2(\mKKg,\ml)-h_5(\mKKg,\ml)-h_8(\mKKg,\ml)\right.
\nonumber\\
&
-\frac{1}{2}[g_2(\mKKs,\ml)+g_5(\mKKs,\ml)+g_8(\mKKs,\ml)-h_2(\mKKs,\ml)+h_5(\mKKs,\ml)-h_8(\mKKs,\ml)]\nonumber\\
&
+\frac{1}{4}[i_2(\mKKg,\ml)+2j_2(\mKKg,\ml,\mKKs)-2s_1(\mKKg,\ml)+n_1(\mKKg,\ml)+2n_{11}(\mKKg,\ml)]\nonumber\\
&
\left.-\frac{1}{2}[-s_8(\mKKg,\ml,\mKKs)+s_{12}(\mKKg,\ml,\mKKs)]\right\}
k_{3\alpha}
\nonumber\\
&
+\left\{3[g_3(\mKKg,\ml)+g_6(\mKKg,\ml)+g_9(\mKKg,\ml)]+h_3(\mKKg,\ml)-h_6(\mKKg,\ml)-h_9(\mKKg,\ml)\right.
\nonumber\\
&
-\frac{1}{2}[g_3(\mKKs,\ml)+g_6(\mKKs,\ml)+g_9(\mKKs,\ml)-h_3(\mKKs,\ml)+h_6(\mKKs,\ml)-h_9(\mKKs,\ml)]\nonumber\\
&
+\frac{1}{4}[i_3(\mKKg,\ml)+2j_3(\mKKg,\ml,\mKKs)-2s_1(\mKKg,\ml)\nonumber\\
&
\left.-\frac{1}{2}[-s_8(\mKKg,\ml,\mKKs)+s_{12}(\mKKg,\ml,\mKKs)]\right\}
k_{4\alpha},\\
f_{2L\alpha}
&
=\left\{-3g_1(\mKKg,\ml)+3g_4(\mKKg,\ml)+2g_7(\mKKg,\ml)+h_1(\mKKg,\ml)-h_4(\mKKg,\ml)-h_7(\mKKg,\ml)\right.
\nonumber\\
&
-\frac{1}{2}[g_1(\mKKs,\ml)-g_4(\mKKs,\ml)+g_7(\mKKs,\ml)-h_1(\mKKs,\ml)+h_4(\mKKs,\ml)-h_7(\mKKs,\ml)]\nonumber\\
&
+\frac{1}{4}[i_4(\mKKg,\ml)-2j_1(\mKKg,\ml,\mKKs)-2j_1(\mKKs,\ml,\mKKg)]\nonumber\\
&
-3[e_1(\mKKg,\ml,\mKKs)-e_1(\mKKs,\ml,\mKKg)]\nonumber\\
&
+\frac{1}{4}[n_2(\mKKg,\ml)+2n_{12}(\mKKg,\ml)+l_1(\ml,\mKKs)-2s_2(\mKKg,\ml)]\nonumber\\
&
\left.+\frac{1}{2}[r_1(\mKKg,\ml,\mKKs)+r_{10}(\mKKg,\ml,\mKKs)-s_8(\mKKg,\ml,\mKKs)+s_{12}(\mKKg,\ml,\mKKs)]\right\}
k_{1\alpha}
\nonumber\\
&
+\left\{-3g_2 (\mKKg,\ml)+ 3 g_5(\mKKg,\ml) + 2 g_8(\mKKg,\ml)+h2 (\mKKg,\ml)- h5(\mKKg,\ml) - h_8(\mKKg,\ml)\right.\nonumber\\
&
-\frac{1}{2}[g_2(\mKKs,\ml) - g_5(\mKKs,\ml) + g_8(\mKKs,\ml) - h_2(\mKKs,\ml) + h_5(\mKKs,\ml) - h_8(\mKKs,\ml)]\nonumber\\
&
+\frac{1}{4}[i_5(\mKKs,\ml)-2j_2(\mKKg,\ml,\mKKs)-2j_2(\mKKs,\ml,\mKKg)\nonumber\\
&
+n_3(\mKKg,\ml)+2n_{13}(\mKKg,\ml)+l_2(\ml,\mKKs)-2s_3(\mKKg,\ml)]\nonumber\\
&
\left.-\frac{1}{2}[-r_2(\mKKg,\ml,\mKKs)-r_{11}(\mKKg,\ml,\mKKs)-s_8(\mKKg,\ml,\mKKs)+s_{12}(\mKKg,\ml,\mKKs)]\right\}
k_{3\alpha}
\nonumber\\
&
+\left\{-3g_3 (\mKKg,\ml)+ 3 g_6(\mKKg,\ml) + 2 g_9(\mKKg,\ml)+h3 (\mKKg,\ml)- h6(\mKKg,\ml) - h_9(\mKKg,\ml)\right.\nonumber\\
&
-\frac{1}{2}[g_3(\mKKs,\ml) - g_6(\mKKs,\ml) + g_9(\mKKs,\ml) - h_3(\mKKs,\ml) + h_6(\mKKs,\ml) - h_9(\mKKs,\ml)]\nonumber\\
&
+\frac{1}{4}[i_6(\mKKs,\ml)-2j_3(\mKKg,\ml,\mKKs)-2j_3(\mKKs,\ml,\mKKg)\nonumber\\
&
-3[e_2(\mKKg,\ml,\mKKs)-e_2(\mKKs,\ml,\mKKg)]\nonumber
\end{align}
\begin{align}
&
+\frac{1}{4}[n_4(\mKKg,\ml)+2n_{14}(\mKKg,\ml)+l_3(\ml,\mKKs)-2s_4(\mKKg,\ml)]
\nonumber\\
&
\left.-\frac{1}{2}[-r_3(\mKKg,\ml,\mKKs)-r_{12}(\mKKg,\ml,\mKKs)-s_8(\mKKg,\ml,\mKKs)+s_{12}(\mKKg,\ml,\mKKs)]\right\}
k_{4\alpha},\\
f_{3L\alpha}
&
=\left\{\frac{1}{4} [g_1(\mKKg,\ml) + 4g_4(\mKKg,\ml)-8g_7(\mKKg,\ml)+ 4 (h_1(\mKKg,\ml) + h_4(\mKKg,\ml) + h_7(\mKKg,\ml))]\right.\nonumber\\
&
-\frac{1}{2} [g_1(\mKKs,\ml) + g_4(\mKKs,\ml)- g_7(\mKKs,\ml)- h_1(\mKKs,\ml)- h_4(\mKKs,\ml) + h_7(\mKKs,\ml)]\nonumber\\
&
+ \frac{1}{4}(i_7(\mKKs,\ml) - 2 j_1(\mKKg,\ml,\mKKs)) - 3 e_1(\mKKg,\ml,\mKKs)\nonumber\\ 
&
+ \frac{1}{4} (n_5(\mKKg,\ml) + 2 n_{12}(\mKKg,\ml))+ \frac{1}{4}[l_4(\ml,\mKKs) -  2 s_2(\mKKg,\ml)]\nonumber\\
&
- \frac{1}{2}[- t_1(\mKKg,\ml,\mKKs) - r_4(\mKKg,\ml,\mKKs) - r_{13}(\mKKg,\ml,\mKKs) \nonumber\\
&
\left.+ s_9(\mKKg,\ml,\mKKs) -  s_{13}(\mKKg,\ml,\mKKs)]\right\}
k_{1\alpha}
\nonumber\\
&
+\left\{\frac{1}{4}[g_2(\mKKg,\ml) + 4 g_5(\mKKg,\ml) - 8 g_8(\mKKg,\ml) + 4 (h_2(\mKKg,\ml) + h_5(\mKKg,\ml) + h_8(\mKKg,\ml))]\right.\nonumber\\
&
-\frac{1}{2}[g_2(\mKKs,\ml)  + g_5(\mKKs,\ml)  - g_8(\mKKs,\ml)  - h_2(\mKKs,\ml) - h_5(\mKKs,\ml) + h_8(\mKKs,\ml)]\nonumber\\
&
+ \frac{1}{4}[i_8(\mKKs,\ml) + 2 j_2(\mKKg,\ml,\mKKs) +n_6(\mKKg,\ml) + 2 n_{13}(\mKKg,\ml)]\nonumber\\
&
+\frac{1}{4}l_5(\ml,\mKKs)-\frac{1}{2}[ -s_3(\mKKg,\ml) - t_2(\mKKg,\ml,\mKKs) -  r_5(\mKKg,\ml,\mKKs) - r_{14}(\mKKg,\ml,\mKKs) \nonumber\\
&
\left.+ s_{10}(\mKKg,\ml,\mKKs) -s_{14}(\mKKg,\ml,\mKKs)]\right\}
k_{3\alpha}
\nonumber\\
&
+\left\{\frac{1}{4}[g_3(\mKKg,\ml)  + 4 g_6(\mKKg,\ml)  - 8 g_9(\mKKg,\ml)  + 4 (h_3(\mKKg,\ml)  + h_6(\mKKg,\ml)  + h_9(\mKKg,\ml) )]\right.\nonumber\\
&
-\frac{1}{2}(g_3(\mKKs,\ml) + g_6(\mKKs,\ml)- g_9(\mKKs,\ml)- h_3(\mKKs,\ml) - h_6(\mKKs,\ml)+ h_9(\mKKs,\ml)) \nonumber\\
&
+ \frac{1}{4}[i_9(\mKKs,\ml) + 2 j_3(\mKKg,\ml,\mKKs)] - 3 e_2(\mKKs,\ml,\mKKg) + \frac{1}{4}[n_7(\mKKg,\ml) + 2 n_{14}(\mKKg,\ml)]\nonumber\\
&
+\frac{1}{4}l_6(\ml,\mKKs)-\frac{1}{2}[-s_4(\mKKg,\ml,\mKKs) -t_3(\mKKg,\ml,\mKKs) -r_6(\mKKg,\ml,\mKKs) - r_{15}(\mKKg,\ml,\mKKs)\nonumber\\
&
\left.+s_{11}(\mKKg,\ml,\mKKs) - s_{15}(\mKKg,\ml,\mKKs)]\right\}
k_{4\alpha}\\
f_{4L\alpha},
&
=\left\{\frac{1}{4} [12 g_1(\mKKg,\ml) - 12 g_4(\mKKg,\ml) + g_7(\mKKg,\ml)\right.\nonumber\\
&+ 4 (-h_1(\mKKg,\ml) + h_4(\mKKg,\ml) + h_7(\mKKg,\ml))]\nonumber\\
&
- \frac{1}{2} (-g_1(\mKKs,\ml) + g_4(\mKKs,\ml)  + g_7(\mKKs,\ml) + h_1(\mKKs,\ml) - h_4(\mKKs,\ml) - h_7(\mKKs,\ml) ) \nonumber\\
&
+\frac{1}{4}[i_{10}(\mKKs,\ml)  - 2 j_1(\mKKg,\ml,\mKKs)] -3e_1(\mKKs,\ml,\mKKg) + \frac{1}{4}[ n_8(\mKKg,\ml) +l_7(\ml,\mKKs)]\nonumber
\end{align}
\begin{align}
&
+\frac{1}{2}[s_5(\mKKg,\ml) + s_2(\mKKg,\ml)]-\frac{1}{2}[-t_4(\mKKg,\ml,\mKKs) - r_7(\mKKg,\ml,\mKKs) - r_{16}(\mKKg,\ml,\mKKs)\nonumber\\
&
\left. - s_8(\mKKg,\ml,\mKKs) +s_{12}(\mKKg,\ml,\mKKs)]\right\}
k_{1\alpha}
\nonumber\\
&
+\left\{\frac{1}{4}[12 g_2(\mKKg,\ml) - 12 g_5(\mKKg,\ml) + g_8(\mKKg,\ml) \right.\nonumber\\
&+ 4 (-h_2(\mKKg,\ml) + h_5 (\mKKg,\ml)+ h_8(\mKKg,\ml))]\nonumber\\
&
 -\frac{1}{2} (-g_2(\mKKs,\ml) + g_5(\mKKs,\ml)+ g_8(\mKKs,\ml) + h_2(\mKKs,\ml)- h_5(\mKKs,\ml)- h_8(\mKKs,\ml))\nonumber\\
&
 +\frac{1}{4}[i_{11}(\mKKs,\ml) + 2 j_2(\mKKg,\ml,\mKKs) + n_9 (\mKKg,\ml)+ l_8(\ml,\mKKs)]\nonumber\\
&
+ \frac{1}{2}[s_6(\mKKg,\ml) +s_3(\mKKg,\ml)]-\frac{1}{2}[- t_5(\mKKg,\ml,\mKKs) - r_8(\mKKg,\ml,\mKKs) -  r_{17}(\mKKg,\ml,\mKKs)\nonumber\\
&
 \left.+ s_8(\mKKg,\ml,\mKKs) -s_{12}(\mKKg,\ml,\mKKs)]\right\}
k_{3\alpha}
\nonumber\\
&
+\left\{\frac{1}{4}[12 g_3(\mKKg,\ml)- 12 g_6(\mKKg,\ml) + g_9(\mKKg,\ml) \right.\nonumber\\
&+ 4 (-h_3(\mKKg,\ml) + h_6(\mKKg,\ml) + h_9(\mKKg,\ml))]\nonumber\\
&
-\frac{1}{2}(-g_3(\mKKs,\ml) + g_6(\mKKs,\ml)+ g_9(\mKKs,\ml)+ h_3(\mKKs,\ml)- h_6(\mKKs,\ml)- h_9(\mKKs,\ml))\nonumber\\
&
+\frac{1}{4}[i_{12}(\mKKs,\ml) + 2  j_3(\mKKg,\ml,\mKKs)]-3e_2(\mKKs,\ml,\mKKg)+\frac{1}{4}[n_{10}(\mKKg,\ml) + l_9(\ml,\mKKs)]\nonumber\\
&
+\frac{1}{2}[s_7(\mKKg,\ml) - s_4(\mKKg,\ml)]-\frac{1}{2}[-t_6(\mKKg,\ml) - r_9(\mKKg,\ml,\mKKs)- r_{18}(\mKKg,\ml,\mKKs)\nonumber\\
&
\left.+s_8 (\mKKg,\ml,\mKKs)- s_{12}(\mKKg,\ml,\mKKs)]\right\}
k_{4\alpha}.
\end{align}
The coefficients of color octet part, $H^{\mu\nu}_i$, are given as
\begin{align}
h_{1R\alpha}
&
=\left\{-2 (g_1(\mKKg,\mr) + g_4(\mKKg,\mr) + g_7(\mKKg,\mr))- \frac{2}{3} (h_1(\mKKg,\mr) - h_4(\mKKg,\mr) - h_7(\mKKg,\mr))\right.\nonumber\\
&
+ \frac{1}{3} (g_1(\mKKs,\mr)+ g_4(\mKKs,\mr) + g_7(\mKKs,\mr) - h_1(\mKKs,\mr)+ h_4(\mKKs,\mr) - h_7(\mKKs,\mr))\nonumber\\
&\left.+ \frac{3}{2}[i_1(\mKKs,\mr) - 2 j_1(\mKKg,\mr,\mKKs)] -18e_1(\mKKs,\mr,\mKKg)\right\}
k_{1\alpha}
\nonumber\\
&
+\left\{-2(g_2(\mKKg,\mr)  + g_5(\mKKg,\mr)  + g_8(\mKKg,\mr) ) - \frac{2}{3}(h_2(\mKKg,\mr)  - h_5(\mKKg,\mr)  - h_8(\mKKg,\mr) )\right.\nonumber\\
&
+ \frac{1}{3}(g_2(\mKKs,\mr)+ g_5(\mKKs,\mr)+ g_8(\mKKs,\mr)- h_2(\mKKs,\mr)+ h_5(\mKKs,\mr)- h_8(\mKKs,\mr))\nonumber\\
&
\left.+ \frac{3}{2}[i_2(\mKKs,\mr)  + 2j_2(\mKKg,\mr,\mKKs)]\right\}
k_{3\alpha}
\nonumber
\end{align}
\begin{align}
&
+\left\{-2(g_3(\mKKg,\mr) + g_6(\mKKg,\mr) + g_9(\mKKg,\mr)) - \frac{2}{3} (h_3(\mKKg,\mr) - h_6(\mKKg,\mr) - h_9(\mKKg,\mr))\right.\nonumber\\
&
+\frac{1}{3}(g_3(\mKKs,\mr)+ g_6(\mKKs,\mr)+ g_9(\mKKs,\mr)- h_3(\mKKs,\mr)+ h_6(\mKKs,\mr) - h_9(\mKKs,\mr))\nonumber\\
&
\left.+ \frac{3}{2}[i_3(\mKKs,\mr) - 2 j_3(\mKKg,\mr,\mKKs)] -18e_2(\mKKs,\mr,\mKKg)\right\}
k_{4\alpha},
\\
h_{2R\alpha}
&
=-\left\{2 g_1(\mKKg,\mr) - 2g_4(\mKKg,\mr) - \frac{4}{3}g_7(\mKKg,\mr) -\frac{2}{3}(h_1(\mKKg,\mr) - h_4(\mKKg,\mr) - h_7(\mKKg,\mr))\right.\nonumber\\
&
+\frac{1}{3}(g_1(\mKKs,\mr)- g_4(\mKKs,\mr)+ g_7(\mKKs,\mr)- h_1(\mKKs,\mr)+ h_4(\mKKs,\mr)- h_7(\mKKs,\mr))\nonumber\\
&
+ \frac{3}{2}[i_4(\mKKs,\mr) - 2j_1(\mKKg,\mr,\mKKs)- 2j_1(\mKKs,\mr,\mKKg)] -18[e_1(\mKKg,\mr,\mKKs)-e_1(\mKKs,\mr,\mKKg)]\nonumber\\
&
+ \left.\frac{3}{2}[l_1(\mr,\mKKs)- 2r_1(\mKKg,\mr,\mKKs)- 2r_{10}(\mKKg,\mr,\mKKs)]\right\}
k_{1\alpha}
\nonumber\\
&
-\left\{2 g_2(\mKKg,\mr) - 2g_5(\mKKg,\mr) - \frac{4}{3}g_8(\mKKg,\mr) -\frac{2}{3}(h_2(\mKKg,\mr) - h_5(\mKKg,\mr) - h_8(\mKKg,\mr))\right.\nonumber\\
&
+\frac{1}{3}(g_2(\mKKs,\mr)- g_5(\mKKs,\mr)+ g_8(\mKKs,\mr)- h_2(\mKKs,\mr)+ h_5(\mKKs,\mr)- h_8(\mKKs,\mr))\nonumber\\
&
+ \frac{3}{2}[i_5(\mKKs,\mr) - 2j_2(\mKKg,\mr,\mKKs)- 2j_2(\mKKs,\mr,\mKKg)]\nonumber\\
&
- \left.3[r_2(\mKKg,\mr,\mKKs)+r_{11}(\mKKg,\mr,\mKKs)]\right\}
k_{3\alpha}
\nonumber\\
&
-\left\{2 g_3(\mKKg,\mr) - 2g_6(\mKKg,\mr) - \frac{4}{3}g_9(\mKKg,\mr) -\frac{2}{3}(h_2(\mKKg,\mr) - h_5(\mKKg,\mr) - h_8(\mKKg,\mr))\right.\nonumber\\
&
+\frac{1}{3}(g_3(\mKKs,\mr)- g_6(\mKKs,\mr)+ g_9(\mKKs,\mr)- h_3(\mKKs,\mr)+ h_6(\mKKs,\mr)- h_9(\mKKs,\mr))\nonumber\\
&
+ \frac{3}{2}[i_6(\mKKs,\mr) - 2j_3(\mKKg,\mr,\mKKs)- 2j_3(\mKKs,\mr,\mKKg)] -18[e_3(\mKKg,\mr,\mKKs)-e_3(\mKKs,\mr,\mKKg)]\nonumber\\
&
+ \left.\frac{3}{2}[l_3(\mr,\mKKs)- 2r_3(\mKKg,\mr,\mKKs)- 2r_{12}(\mKKg,\mr,\mKKs)]\right\}
k_{4\alpha},
\end{align}
\begin{align}
h_{3R\alpha}
&
=-\frac{1}{6}(g_1(\mKKg,\mr) + 4 g_4(\mKKg,\mr) - 8 g_7(\mKKg,\mr) + 4 (h_1(\mKKg,\mr) + h_4(\mKKg,\mr) + h_7(\mKKg,\mr)))\nonumber\\
&
+\left\{\frac{1}{3}g_1(\mKKs,\mr) + g_4(\mKKs,\mr)- g_7(\mKKs,\mr)- h_1(\mKKs,\mr)- h_4(\mKKs,\mr)+ h_7(\mKKs,\mr))\right.\nonumber\\
&
+\frac{3}{2}[i_7(\mKKs,\mr) - 2j_1(\mKKg,\mr,\mKKs)] -18 e_1(\mKKg,\mr,\mKKs) + \frac{3}{2}l_4(\mr,\mKKs)\nonumber\\
&
\left.-3 ( t_1(\mKKs,\mr) + r_4(\mKKg,\mr,\mKKs)+r_{13}(\mKKg,\mr,\mKKs))\right\}
k_{1\alpha}
\nonumber\\
&
-\frac{1}{6}(g_2(\mKKg,\mr) + 4 g_5(\mKKg,\mr) - 8 g_8(\mKKg,\mr) + 4 (h_2(\mKKg,\mr) + h_5(\mKKg,\mr) + h_8(\mKKg,\mr)))\nonumber\\
&
+\left\{\frac{1}{3}g_2(\mKKs,\mr) + g_5(\mKKs,\mr)- g_8(\mKKs,\mr)- h_2(\mKKs,\mr)- h_5(\mKKs,\mr)+ h_8(\mKKs,\mr))\right.\nonumber
\end{align}
\begin{align}
&
+\frac{3}{2}[i_8(\mKKs,\mr) + 2j_2(\mKKg,\mr,\mKKs)] + \frac{3}{2}l_5(\mr,\mKKs)\nonumber\\
&\left.-3 ( t_2(\mKKs,\mr) + r_5(\mKKg,\mr,\mKKs)+r_{14}(\mKKg,\mr,\mKKs))\right\}
k_{3\alpha}
\nonumber\\
&
-\frac{1}{6}(g_3(\mKKg,\mr) + 4 g_6(\mKKg,\mr) - 8 g_9(\mKKg,\mr) + 4 (h_3(\mKKg,\mr) + h_6(\mKKg,\mr) + h_9(\mKKg,\mr)))\nonumber\\
&
+\left\{\frac{1}{3}g_3(\mKKs,\mr) + g_6(\mKKs,\mr)- g_9(\mKKs,\mr)- h_4(\mKKs,\mr)- h_6(\mKKs,\mr)+ h_9(\mKKs,\mr))\right.\nonumber\\
&
+\frac{3}{2}[i_9(\mKKs,\mr) - 2j_3(\mKKg,\mr,\mKKs)] -18 e_3(\mKKg,\mr,\mKKs) + \frac{3}{2}l_6(\mr,\mKKs)\nonumber\\
&
\left.-3 ( t_3(\mKKs,\mr) + r_6(\mKKg,\mr,\mKKs)+r_{15}(\mKKg,\mr,\mKKs))\right\}
k_{4\alpha},
\\
h_{4R\alpha}
&
=-\left\{\frac{1}{6}(12 g_1(\mKKg,\mr)  - 12 g_4(\mKKg,\mr)  + g_7(\mKKg,\mr)\right. \nonumber\\
&
+ 4 (-h_1(\mKKg,\mr)  + h_4 (\mKKg,\mr) + h_7(\mKKg,\mr) ))\nonumber\\
&
+\frac{1}{3}(-g_1(\mKKs,\mr)+ g_4(\mKKs,\mr)+ g_7(\mKKs,\mr)+ h_1(\mKKs,\mr)- h_4(\mKKs,\mr)- h_7(\mKKs,\mr))\nonumber\\
&+\frac{3}{2}[i_{10}(\mKKs,\mr) - 2j_1(\mKKg,\mr,\mKKs)] - 18 e_1(\mKKg,\mr,\mKKs) +\frac{3}{2} l_7(\mr,\mKKs)\nonumber\\
&
-3\left.(t_4(\mKKs,\mr)+r_7(\mKKg,\mr,\mKKs)+r_{16}(\mKKg,\mr,\mKKs))\right\}
k_{1\alpha}
\nonumber\\
&
-\left\{\frac{1}{6}(12 g_2(\mKKg,\mr)  - 12 g_5(\mKKg,\mr)  + g_8(\mKKg,\mr)\right.\nonumber\\
&
+ 4 (-h_2(\mKKg,\mr)  + h_5(\mKKg,\mr) + h_8(\mKKg,\mr) ))\nonumber\\
&
+\frac{1}{3}(-g_2(\mKKs,\mr)+ g_5(\mKKs,\mr)+ g_8(\mKKs,\mr)+ h_2(\mKKs,\mr)- h_5(\mKKs,\mr)- h_8(\mKKs,\mr))\nonumber\\
&
+\frac{3}{2}[i_{11}(\mKKs,\mr) - 2j_2(\mKKg,\mr,\mKKs)] +\frac{3}{2} l_8(\mr,\mKKs)\nonumber\\
&
-3\left.(t_5(\mKKs,\mr)+r_8(\mKKg,\mr,\mKKs)+r_{17}(\mKKg,\mr,\mKKs))\right\}
k_{3\alpha}
\nonumber\\
&
-\left\{\frac{1}{6}(12 g_3(\mKKg,\mr)  - 12 g_6(\mKKg,\mr)  + g_9(\mKKg,\mr)\right. \nonumber\\
&
+ 4 (-h_3(\mKKg,\mr)  + h_6(\mKKg,\mr) + h_9(\mKKg,\mr) ))\nonumber\\
&
+\frac{1}{3}(-g_3(\mKKs,\mr)+ g_6(\mKKs,\mr)+ g_9(\mKKs,\mr)+ h_3(\mKKs,\mr)- h_6(\mKKs,\mr)- h_9(\mKKs,\mr))\nonumber\\
&
+\frac{3}{2}[i_{12}(\mKKs,\mr) - 2j_3(\mKKg,\mr,\mKKs)] - 18 e_3(\mKKg,\mr,\mKKs) +\frac{3}{2} l_9(\mr,\mKKs)\nonumber\\
&
-3\left.(t_6(\mKKs,\mr)+r_9(\mKKg,\mr,\mKKs)+r_{18}(\mKKg,\mr,\mKKs))\right\}
k_{4\alpha},\\
h_{1L\alpha}
&
=h_{1R\alpha}
\end{align}
\begin{align}
h_{2L\alpha}
&
=-\left\{2 g_1(\mKKg,\ml) - 2g_4(\mKKg,\ml) - \frac{4}{3}g_7(\mKKg,\ml) -\frac{2}{3}(h_1(\mKKg,\ml) - h_4(\mKKg,\ml) - h_7(\mKKg,\ml))\right.\nonumber\\
&
+\frac{1}{3}(g_1(\mKKs,\ml)- g_4(\mKKs,\ml)+ g_7(\mKKs,\ml)- h_1(\mKKs,\ml)+ h_4(\mKKs,\ml)- h_7(\mKKs,\ml))\nonumber\\
&
+ \frac{3}{2}[i_4(\mKKs,\ml) - 2j_1(\mKKg,\ml,\mKKs)- 2j_1(\mKKs,\ml,\mKKg)] -18[e_1(\mKKg,\ml,\mKKs)-e_1(\mKKs,\ml,\mKKg)]\nonumber\\
&
+ \left.\frac{3}{2}[l_1(\ml,\mKKs)+ 2r_1(\mKKg,\ml,\mKKs)+ 2r_{10}(\mKKg,\ml,\mKKs)]\right\}
k_{1\alpha}
\nonumber\\
&
-\left\{2 g_2(\mKKg,\ml) - 2g_5(\mKKg,\ml) - \frac{4}{3}g_8(\mKKg,\ml) -\frac{2}{3}(h_2(\mKKg,\ml) - h_5(\mKKg,\ml) - h_8(\mKKg,\ml))\right.\nonumber\\
&
+\frac{1}{3}(g_2(\mKKs,\ml)- g_5(\mKKs,\ml)+ g_8(\mKKs,\ml)- h_2(\mKKs,\ml)+ h_5(\mKKs,\ml)- h_8(\mKKs,\ml))\nonumber\\
&
+ \frac{3}{2}[i_5(\mKKs,\ml) - 2j_2(\mKKg,\ml,\mKKs)- 2j_2(\mKKs,\ml,\mKKg)]\nonumber\\
&
+ \left.3[r_2(\mKKg,\ml,\mKKs)+r_{11}(\mKKg,\ml,\mKKs)]\right\}
k_{3\alpha}
\nonumber\\
&
-\left\{2 g_3(\mKKg,\ml) - 2g_6(\mKKg,\ml) - \frac{4}{3}g_9(\mKKg,\ml) -\frac{2}{3}(h_2(\mKKg,\ml) - h_5(\mKKg,\ml) - h_8(\mKKg,\ml))\right.\nonumber\\
&
+\frac{1}{3}(g_3(\mKKs,\ml)- g_6(\mKKs,\ml)+ g_9(\mKKs,\ml)- h_3(\mKKs,\ml)+ h_6(\mKKs,\ml)- h_9(\mKKs,\ml))\nonumber\\
&
+ \frac{3}{2}[i_6(\mKKs,\ml) - 2j_3(\mKKg,\ml,\mKKs)- 2j_3(\mKKs,\ml,\mKKg)] -18[e_3(\mKKg,\ml,\mKKs)-e_3(\mKKs,\ml,\mKKg)]\nonumber\\
&
+ \left.\frac{3}{2}[l_3(\ml,\mKKs)+2r_3(\mKKg,\ml,\mKKs)+2r_{12}(\mKKg,\ml,\mKKs)]\right\}
k_{4\alpha},\\
h_{3L\alpha}
&
=-\frac{1}{6}(g_1(\mKKg,\ml) + 4 g_4(\mKKg,\ml) - 8 g_7(\mKKg,\ml) + 4 (h_1(\mKKg,\ml) + h_4(\mKKg,\ml) + h_7(\mKKg,\ml)))\nonumber\\
&
+\left\{\frac{1}{3}g_1(\mKKs,\ml) + g_4(\mKKs,\ml)- g_7(\mKKs,\ml)- h_1(\mKKs,\ml)- h_4(\mKKs,\ml)+ h_7(\mKKs,\ml))\right.\nonumber\\
&
+\frac{3}{2}[i_7(\mKKs,\ml) - 2j_1(\mKKg,\ml,\mKKs)] -18 e_1(\mKKg,\ml,\mKKs) + \frac{3}{2}l_4(\ml,\mKKs)\nonumber\\
&
\left.+3 ( t_1(\mKKs,\ml) + r_4(\mKKg,\ml,\mKKs)+r_{13}(\mKKg,\ml,\mKKs))\right\}
k_{1\alpha}
\nonumber\\
&
-\frac{1}{6}(g_2(\mKKg,\ml) + 4 g_5(\mKKg,\ml) - 8 g_8(\mKKg,\ml) + 4 (h_2(\mKKg,\ml) + h_5(\mKKg,\ml) + h_8(\mKKg,\ml)))\nonumber\\
&
+\left\{\frac{1}{3}g_2(\mKKs,\ml) + g_5(\mKKs,\ml)- g_8(\mKKs,\ml)- h_2(\mKKs,\ml)- h_5(\mKKs,\ml)+ h_8(\mKKs,\ml))\right.\nonumber\\
&
+\frac{3}{2}[i_8(\mKKs,\ml) + 2j_2(\mKKg,\ml,\mKKs)] + \frac{3}{2}l_5(\ml,\mKKs)\nonumber\\
&
\left.+3 ( t_2(\mKKs,\ml) + r_5(\mKKg,\ml,\mKKs)+r_{14}(\mKKg,\ml,\mKKs))\right\}
k_{3\alpha}
\nonumber
\end{align}
\begin{align}
&
-\frac{1}{6}(g_3(\mKKg,\ml) + 4 g_6(\mKKg,\ml) - 8 g_9(\mKKg,\ml) + 4 (h_3(\mKKg,\ml) + h_6(\mKKg,\ml) + h_9(\mKKg,\ml)))\nonumber\\
&
+\left\{\frac{1}{3}g_3(\mKKs,\ml) + g_6(\mKKs,\ml)- g_9(\mKKs,\ml)- h_4(\mKKs,\ml)- h_6(\mKKs,\ml)+ h_9(\mKKs,\ml))\right.\nonumber\\
&
+\frac{3}{2}[i_9(\mKKs,\ml) - 2j_3(\mKKg,\ml,\mKKs)] -18 e_3(\mKKg,\ml,\mKKs) + \frac{3}{2}l_6(\ml,\mKKs)\nonumber\\
&
\left.+3 ( t_3(\mKKs,\ml) + r_6(\mKKg,\ml,\mKKs)+r_{15}(\mKKg,\ml,\mKKs))\right\}
k_{4\alpha},\\
h_{4L\alpha}
&
=-\left\{\frac{1}{6}(12 g_1(\mKKg,\ml)  - 12 g_4(\mKKg,\ml)  + g_7(\mKKg,\ml)\right. \nonumber\\
&
+ 4 (-h_1(\mKKg,\ml)  + h_4 (\mKKg,\ml) + h_7(\mKKg,\ml) ))\nonumber\\
&
+\frac{1}{3}(-g_1(\mKKs,\ml)+ g_4(\mKKs,\ml)+ g_7(\mKKs,\ml)+ h_1(\mKKs,\ml)- h_4(\mKKs,\ml)- h_7(\mKKs,\ml))\nonumber\\
&
+\frac{3}{2}[i_{10}(\mKKs,\ml) - 2j_1(\mKKg,\ml,\mKKs)] - 18 e_1(\mKKg,\ml,\mKKs) +\frac{3}{2} l_7(\ml,\mKKs)\nonumber\\
&
+3\left.(t_4(\mKKs,\ml)+r_7(\mKKg,\ml,\mKKs)+r_{16}(\mKKg,\ml,\mKKs))\right\}
k_{1\alpha}
\nonumber\\
&
-\left\{\frac{1}{6}(12 g_2(\mKKg,\ml)  - 12 g_5(\mKKg,\ml)  + g_8(\mKKg,\ml)\right.\nonumber\\
&
+ 4 (-h_2(\mKKg,\ml)  + h_5(\mKKg,\ml) + h_8(\mKKg,\ml) ))\nonumber\\
&
+\frac{1}{3}(-g_2(\mKKs,\ml)+ g_5(\mKKs,\ml)+ g_8(\mKKs,\ml)+ h_2(\mKKs,\ml)- h_5(\mKKs,\ml)- h_8(\mKKs,\ml))\nonumber\\
&
+\frac{3}{2}[i_{11}(\mKKs,\ml) - 2j_2(\mKKg,\ml,\mKKs)] +\frac{3}{2} l_8(\ml,\mKKs)\nonumber\\
&
+3\left.(t_5(\mKKs,\ml)+r_8(\mKKg,\ml,\mKKs)+r_{17}(\mKKg,\ml,\mKKs))\right\}
k_{3\alpha}
\nonumber\\
&
-\left\{\frac{1}{6}(12 g_3(\mKKg,\ml)  - 12 g_6(\mKKg,\ml)  + g_9(\mKKg,\ml)\right. \nonumber\\
&
+ 4 (-h_3(\mKKg,\ml)  + h_6(\mKKg,\ml) + h_9(\mKKg,\ml) ))\nonumber\\
&
+\frac{1}{3}(-g_3(\mKKs,\ml)+ g_6(\mKKs,\ml)+ g_9(\mKKs,\ml)+ h_3(\mKKs,\ml)- h_6(\mKKs,\ml)- h_9(\mKKs,\ml))\nonumber\\
&
+\frac{3}{2}[i_{12}(\mKKs,\ml) - 2j_3(\mKKg,\ml,\mKKs)] - 18 e_3(\mKKg,\ml,\mKKs) +\frac{3}{2} l_9(\ml,\mKKs)\nonumber\\
&
+3\left.(t_6(\mKKs,\ml)+r_9(\mKKg,\ml,\mKKs)+r_{18}(\mKKg,\ml,\mKKs))\right\}
k_{4\alpha}.
\end{align}
These coefficients, $g(m_a,m_b), h(m_a,m_b), \cdots$, are written in terms of Feynman integral as follows;
\begin{align}
g_1(m_a,m_b)&=\int^1_0dx\int^1_0dy\int^1_0dz\frac{6 y z^2-6 x y^2 z^3}{z \left(m_b^2-m_a^2\right)+m_a^2},\\
g_2(m_a,m_b)&=g_8(m_a,m_b)\nonumber\\
&=\int^1_0dx\int^1_0dy\int^1_0dz\frac{6 y^2 z^3-6 y z^3}{z \left(m_b^2-m_a^2\right)+m_a^2},\\
g_3(m_a,m_b)&=\int^1_0dx\int^1_0dy\int^1_0dz\frac{6 x y^2 z^3-6 y z^3}{z \left(m_b^2-m_a^2\right)+m_a^2},
\end{align}
\begin{align}
g_4(m_a,m_b)&=g_7(m_a,m_b)\nonumber\\
&=\int^1_0dx\int^1_0dy\int^1_0dz-\frac{6 x y^2 z^3}{z \left(m_b^2-m_a^2\right)+m_a^2},\\
g_5(m_a,m_b)&=\int^1_0dx\int^1_0dy\int^1_0dz\frac{6 y^2 z^3-6 y z^3+6 y z^2}{z \left(m_b^2-m_a^2\right)+m_a^2},\\
g_6(m_a,m_b)&=g_9(m_a,m_b)\nonumber\\
&=\int^1_0dx\int^1_0dy\int^1_0dz\frac{6 y^2 z^3-6 y z^3+6 y z^2}{z \left(m_b^2-m_a^2\right)+m_a^2},\\
h_1(m_a,m_b)&=\int^1_0dx\int^1_0dy\int^1_0dz\frac{m_b^2 \left(6 y z^2-6 x y^2 z^3\right)}{\left(z
   \left(m_b^2-m_a^2\right)+m_a^2\right){}^2},\\
h_2(m_a,m_b)&=h_8(m_a,m_b)\nonumber\\
&=\int^1_0dx\int^1_0dy\int^1_0dz\frac{m_b^2 \left(6 y z^2-6 x y^2 z^3\right)}{\left(z
   \left(m_b^2-m_a^2\right)+m_a^2\right){}^2},\\
h_3(m_a,m_b)&=\int^1_0dx\int^1_0dy\int^1_0dz\frac{m_b^2 \left(6 x y^2 z^3-6 y z^3\right)}{\left(z
   \left(m_b^2-m_a^2\right)+m_a^2\right){}^2},\\
h_4(m_a,m_b)&=h_7(m_a,m_b)\nonumber\\
&=\int^1_0dx\int^1_0dy\int^1_0dz-\frac{6 x y^2 z^3 m_b^2}{\left(z \left(m_b^2-m_a^2\right)+m_a^2\right){}^2},\\
h_5(m_a,m_b)&=\int^1_0dx\int^1_0dy\int^1_0dz\frac{m_b^2 \left(6 y^2 z^3-6 y z^3+6 y z^2\right)}{\left(z
   \left(m_b^2-m_a^2\right)+m_a^2\right){}^2},\\
h_6(m_a,m_b)&=h_9(m_a,m_b)\\
&=\frac{m_b^2 \left(6 x y^2 z^3-6 y z^3+6 y z^2\right)}{\left(z
   \left(m_b^2-m_a^2\right)+m_a^2\right){}^2},\\
i_{1}(m_a,m_b)&=i_{4}(m_a,m_b)\nonumber\\
&=\int^1_0dx\int^1_0dy\int^1_0dz\frac{36 x y^2 z^3+18 y z^2}{z \left(m_a^2-m_b^2\right)+m_b^2},\\
i_{2}(m_a,m_b)&=i_{5}(m_a,m_b)\nonumber\\
&=\int^1_0dx\int^1_0dy\int^1_0dz\frac{36 \left(y z^3-y^2 z^3\right)-48 y z^2}{z \left(m_a^2-m_b^2\right)+m_b^2},\\
i_{3}(m_a,m_b)&=i_{6}(m_a,m_b)\nonumber\\
&=\int^1_0dx\int^1_0dy\int^1_0dz\frac{36 \left(y z^3-x y^2 z^3\right)-24 y z^2}{z \left(m_a^2-m_b^2\right)+m_b^2},\\
i_{7}(m_a,m_b)&=\int^1_0dx\int^1_0dy\int^1_0dz\frac{6 y z^2-72 x y^2 z^3}{z \left(m_a^2-m_b^2\right)+m_b^2},\\
i_{8}(m_a,m_b)&=\int^1_0dx\int^1_0dy\int^1_0dz\frac{72 \left(y^2 z^3-y z^3\right)+12 y z^2}{z \left(m_a^2-m_b^2\right)+m_b^2},\\
i_{9}(m_a,m_b)&=\int^1_0dx\int^1_0dy\int^1_0dz\frac{72 \left(x y^2 z^3-y z^3\right)}{z \left(m_a^2-m_b^2\right)+m_b^2},
\end{align}
\begin{align}
i_{10}(m_a,m_b)&=\int^1_0dx\int^1_0dy\int^1_0dz\frac{-156 x y^2 z^3-18 y z^3}{z \left(m_a^2-m_b^2\right)+m_b^2},\\
i_{11}(m_a,m_b)&=\int^1_0dx\int^1_0dy\int^1_0dz\frac{156 \left(y^2 z^3-y z^3\right)+84 y z^2}{z \left(m_a^2-m_b^2\right)+m_b^2},\\
i_{12}(m_a,m_b)&=\int^1_0dx\int^1_0dy\int^1_0dz\frac{156 \left(x y^2 z^3-y z^3\right)+132 y z^2}{z \left(m_a^2-m_b^2\right)+m_b^2},\\
j_{1}(m_a,m_b,m_c)&=\int^1_0dx\int^1_0dy\int^1_0dz\frac{6 x y^2 z^3 m_b^2}{\left(z \left(m_c^2-m_a^2\right)+m_a^2+y z
   \left(m_b^2-m_c^2\right)\right){}^2},\\
j_{2}(m_a,m_b,m_c)&=\int^1_0dx\int^1_0dy\int^1_0dz\frac{m_b^2 \left(6 y z^3-6 y^2 z^3\right)}{\left(z \left(m_c^2-m_a^2\right)+m_a^2+y z
   \left(m_b^2-m_c^2\right)\right){}^2},\\
j_{3}(m_a,m_b,m_c)&=\int^1_0dx\int^1_0dy\int^1_0dz\frac{m_b^2 \left(6 y z^3-6 x y^2 z^3\right)}{\left(z \left(m_c^2-m_a^2\right)+m_a^2+y
   z \left(m_b^2-m_c^2\right)\right){}^2},\\
e_{1}(m_a,m_b)&=\int^1_0dx\int^1_0dy\int^1_0dz\frac{2 x y^2}{y \left(m_a^2-m_b^2\right)+m_b^2},\\
e_{2}(m_a,m_b)&=\int^1_0dx\int^1_0dy\int^1_0dz\frac{2 y^2-2 x y^2}{y \left(m_a^2-m_b^2\right)+m_b^2},\\
n_{1}(m_a,m_b)&=\int^1_0dx\int^1_0dy\int^1_0dz-\frac{12 y z^2}{y z \left(m_a^2-m_b^2\right)+m_b^2},\\
n_{2}(m_a,m_b)&=n_{8}(m_a,m_b)\nonumber\\
&=\int^1_0dx\int^1_0dy\int^1_0dz\frac{6 y (4-4 z) z^2}{y z \left(m_a^2-m_b^2\right)+m_b^2},\\
n_{3}(m_a,m_b)&=n_{9}(m_a,m_b)=0,\\
n_{4}(m_a,m_b)&=n_{10}(m_a,m_b)\nonumber\\
&=\int^1_0dx\int^1_0dy\int^1_0dz\frac{24 y z^2 (z-y z)}{y z \left(m_a^2-m_b^2\right)+m_b^2},\\
n_{5}(m_a,m_b)&=\int^1_0dx\int^1_0dy\int^1_0dz\frac{6 y z^2 (8 z+4)}{y z \left(m_a^2-m_b^2\right)+m_b^2},\\
n_{6}(m_a,m_b)&=\int^1_0dx\int^1_0dy\int^1_0dz\frac{48 y z^2 (-x y z+y z-z)-12 y z^2}{y z \left(m_a^2-m_b^2\right)+m_b^2},\\
n_{7}(m_a,m_b)&=\int^1_0dx\int^1_0dy\int^1_0dz\frac{48 y z^2 (z-y z)}{y z \left(m_a^2-m_b^2\right)+m_b^2},\\
n_{11}(m_a,m_b)&=\int^1_0dx\int^1_0dy\int^1_0dz-\frac{6 y z^2 m_b^2}{\left(y z \left(m_a^2-m_b^2\right)+m_b^2\right){}^2},\\
n_{12}(m_a,m_b)&=\int^1_0dx\int^1_0dy\int^1_0dz\frac{12 y (1-z) z^2 m_b^2}{\left(y z \left(m_a^2-m_b^2\right)+m_b^2\right){}^2},\\
n_{13}(m_a,m_b)&=0,
\end{align}
\begin{align}
n_{14}(m_a,m_b)&=\int^1_0dx\int^1_0dy\int^1_0dz\frac{12 y z^2 m_b^2 (z-y z)}{\left(y z \left(m_a^2-m_b^2\right)+m_b^2\right){}^2},\\
t_{1}(m_a,m_b,m_c)&=\int^1_0dx\int^1_0dy\int^1_0dz\frac{6 y z^2 m_a m_b (1-2 x y z)}{\left(x y z \left(m_c^2-m_a^2\right)+y z
   \left(m_a^2-m_c^2\right)+z \left(m_c^2-m_b^2\right)+m_b^2\right){}^2},\\
t_{2}(m_a,m_b,m_c)&=\int^1_0dx\int^1_0dy\int^1_0dz\frac{6 y z^2 m_a m_b (1-2 x y z)}{\left(x y z \left(m_c^2-m_a^2\right)+y z
   \left(m_a^2-m_c^2\right)+z \left(m_c^2-m_b^2\right)+m_b^2\right){}^2},\\
t_{3}(m_a,m_b,m_c)&=\int^1_0dx\int^1_0dy\int^1_0dz\frac{6 y z^2 m_a m_b (2 (x y z-z)+1)}{\left(x y z \left(m_c^2-m_a^2\right)+y z
   \left(m_a^2-m_c^2\right)+z \left(m_c^2-m_b^2\right)+m_b^2\right){}^2},\\
t_{4}(m_a,m_b,m_c)&=\int^1_0dx\int^1_0dy\int^1_0dz\frac{12 x y^2 z^3 m_a m_b}{\left(z \left(m_a^2-m_b^2\right)+x y z
   \left(m_a^2-m_c^2\right)+y z \left(m_c^2-m_a^2\right)+m_b^2\right){}^2},\\
t_{5}(m_a,m_b,m_c)&=\int^1_0dx\int^1_0dy\int^1_0dz\frac{6 y z^2 m_a m_b (2 (y z-z)+1)}{\left(z \left(m_a^2-m_b^2\right)+x y z
   \left(m_a^2-m_c^2\right)+y z \left(m_c^2-m_a^2\right)+m_b^2\right){}^2},\\
t_{6}(m_a,m_b,m_c)&=\int^1_0dx\int^1_0dy\int^1_0dz\frac{12 y z^2 m_a m_b (x y z-z+1)}{\left(z \left(m_a^2-m_b^2\right)+x y z
   \left(m_a^2-m_c^2\right)+y z \left(m_c^2-m_a^2\right)+m_b^2\right){}^2},\\
l_{1}(m_b,m_c)&=l_{7}(m_b,m_c)\nonumber\\
&=\int^1_0dx\int^1_0dy\int^1_0dz\frac{24 x y^2 z^3}{z \left(m_c^2-m_b^2\right)+m_b^2},\\
l_{2}(m_b,m_c)&=\int^1_0dx\int^1_0dy\int^1_0dz\frac{6 y z^2 (4 z-4 y z)}{z \left(m_c^2-m_b^2\right)+m_b^2},\\
l_{3}(m_b,m_c)&=\int^1_0dx\int^1_0dy\int^1_0dz\frac{6 y z^2 (4 z-4 x y z)}{z \left(m_c^2-m_b^2\right)+m_b^2},\\
l_{4}(m_b,m_c)&=\int^1_0dx\int^1_0dy\int^1_0dz\frac{6 y z^2 (4 x y z-2)}{z \left(m_c^2-m_b^2\right)+m_b^2},\\
l_{5}(m_b,m_c)&=l_{8}(m_b,m_c)\nonumber\\
&=\int^1_0dx\int^1_0dy\int^1_0dz\frac{6 y z^2 (-4 y z+4 z-2)}{z \left(m_c^2-m_b^2\right)+m_b^2},\\
l_{6}(m_b,m_c)&=\int^1_0dx\int^1_0dy\int^1_0dz\frac{6 y z^2 (-4 x y z+4 z-2)}{z \left(m_c^2-m_b^2\right)+m_b^2},\\
l_{9}(m_b,m_c)&=\int^1_0dx\int^1_0dy\int^1_0dz\frac{6 y z^2 (-4 x y z+4 z-4)}{z \left(m_c^2-m_b^2\right)+m_b^2},\\
r_{1}(m_a,m_b,m_c)&=r_{4}(m_a,m_b,m_c)\nonumber\\
&=\int^1_0dx\int^1_0dy\int^1_0dz-\frac{6 x y^2 z^3 m_a m_b}{\left(z \left(m_a^2-m_b^2\right)+x y z
   \left(m_c^2-m_a^2\right)+m_b^2\right){}^2},
\end{align}
\begin{align}
r_{2}(m_a,m_b,m_c)&=\int^1_0dx\int^1_0dy\int^1_0dz\frac{6 y z^2 m_a m_b (y z-z+1)}{\left(z \left(m_a^2-m_b^2\right)+x y z
   \left(m_c^2-m_a^2\right)+m_b^2\right){}^2},\\
r_{3}(m_a,m_b,m_c)&=\int^1_0dx\int^1_0dy\int^1_0dz\frac{6 y z^2 m_a m_b (x y z-z+1)}{\left(z \left(m_a^2-m_b^2\right)+x y z
   \left(m_c^2-m_a^2\right)+m_b^2\right){}^2},\\
r_{5}(m_a,m_b,m_c)&=\int^1_0dx\int^1_0dy\int^1_0dz\frac{6 y z^2 m_a m_b (y z-z)}{\left(z \left(m_a^2-m_b^2\right)+x y z
   \left(m_c^2-m_a^2\right)+m_b^2\right){}^2},\\
r_{6}(m_a,m_b,m_c)&=\int^1_0dx\int^1_0dy\int^1_0dz\frac{6 y z^2 m_a m_b (x y z-z+1)}{\left(z \left(m_a^2-m_b^2\right)+x y z
   \left(m_c^2-m_a^2\right)+m_b^2\right){}^2},\\
r_{7}(m_a,m_b,m_c)&=-2r_{1}(m_a,m_b,m_c),\\
r_{8}(m_a,m_b,m_c)&=\int^1_0dx\int^1_0dy\int^1_0dz\frac{6 y z^2 m_a m_b (2 y z-2 z-1)}{\left(z \left(m_a^2-m_b^2\right)+x y z
   \left(m_c^2-m_a^2\right)+m_b^2\right){}^2},\\
r_{9}(m_a,m_b,m_c)&=\int^1_0dx\int^1_0dy\int^1_0dz\frac{12 y z^2 m_a m_b (x y z-z-1)}{\left(z \left(m_a^2-m_b^2\right)+x y z
   \left(m_c^2-m_a^2\right)+m_b^2\right){}^2},\\
r_{10}(m_a,m_b,m_c)&=\int^1_0dx\int^1_0dy\int^1_0dz\frac{6 x y^2 z^3 m_a m_b}{\left(z \left(m_a^2-m_b^2\right)+x y z
   \left(m_a^2-m_c^2\right)+y z \left(m_c^2-m_a^2\right)+m_b^2\right){}^2},\\
r_{11}(m_a,m_b,m_c)&=\int^1_0dx\int^1_0dy\int^1_0dz\frac{6 y z^2 m_a m_b (y z-z-1)}{\left(z \left(m_a^2-m_b^2\right)+x y z
   \left(m_a^2-m_c^2\right)+y z \left(m_c^2-m_a^2\right)+m_b^2\right){}^2},\\
r_{12}(m_a,m_b,m_c)&=\int^1_0dx\int^1_0dy\int^1_0dz\frac{6 y z^2 m_a m_b (x y z-z-1)}{\left(z \left(m_a^2-m_b^2\right)+x y z
   \left(m_a^2-m_c^2\right)+y z \left(m_c^2-m_a^2\right)+m_b^2\right){}^2},\\
r_{13}(m_a,m_b,m_c)&=\int^1_0dx\int^1_0dy\int^1_0dz\frac{6 y z^2 m_a m_b (1-2 x y z)}{\left(z \left(m_a^2-m_b^2\right)+x y z
   \left(m_a^2-m_c^2\right)+y z \left(m_c^2-m_a^2\right)+m_b^2\right){}^2},\\
r_{14}(m_a,m_b,m_c)&=\int^1_0dx\int^1_0dy\int^1_0dz\frac{6 y z^2 m_a m_b (2 y z-2 z+1)}{\left(z \left(m_a^2-m_b^2\right)+x y z
   \left(m_a^2-m_c^2\right)+y z \left(m_c^2-m_a^2\right)+m_b^2\right){}^2},\\
r_{15}(m_a,m_b,m_c)&=\int^1_0dx\int^1_0dy\int^1_0dz\frac{6 y z^2 m_a m_b (2 x y z-2 z+1)}{\left(z \left(m_a^2-m_b^2\right)+x y z
   \left(m_a^2-m_c^2\right)+y z \left(m_c^2-m_a^2\right)+m_b^2\right){}^2},\\
r_{16}(m_a,m_b,m_c)&=\int^1_0dx\int^1_0dy\int^1_0dz\frac{6 y z^2 m_a m_b (x y z-1)}{\left(z \left(m_a^2-m_b^2\right)+x y z
   \left(m_a^2-m_c^2\right)+y z \left(m_c^2-m_a^2\right)+m_b^2\right){}^2},\\
r_{17}(m_a,m_b,m_c)&=\int^1_0dx\int^1_0dy\int^1_0dz\frac{6 y z^2 m_a m_b (y z-z)}{\left(z \left(m_a^2-m_b^2\right)+x y z
   \left(m_a^2-m_c^2\right)+y z \left(m_c^2-m_a^2\right)+m_b^2\right){}^2},\\
r_{18}(m_a,m_b,m_c)&=\int^1_0dx\int^1_0dy\int^1_0dz\frac{6 y z^2 m_a m_b (x y z-z)}{\left(z \left(m_a^2-m_b^2\right)+x y z
   \left(m_a^2-m_c^2\right)+y z \left(m_c^2-m_a^2\right)+m_b^2\right){}^2},
\end{align}
\begin{align}
s_{1}(m_b,m_c)&=\int^1_0dx\int^1_0dy\int^1_0dz\frac{6 y z^2}{y z \left(m_c^2-m_b^2\right)+m_b^2},\\
s_{2}(m_b,m_c)&=\int^1_0dx\int^1_0dy\int^1_0dz\frac{6 y z^2 (-4 y z+2 z+1)}{y z \left(m_c^2-m_b^2\right)+m_b^2},\\
s_{3}(m_b,m_c)&=\int^1_0dx\int^1_0dy\int^1_0dz\frac{6 y z^2 (2 x y z+2 y z-2 z-1)}{y z \left(m_c^2-m_b^2\right)+m_b^2},\\
s_{4}(m_b,m_c)&=\int^1_0dx\int^1_0dy\int^1_0dz\frac{6 y z^2 (2 y z-2 z-1)}{y z \left(m_c^2-m_b^2\right)+m_b^2},\\
s_{5}(m_b,m_c)&=\int^1_0dx\int^1_0dy\int^1_0dz6 y z^2 (-4 y z+2 z+2) \nonumber\\
&\cdot \left(\frac{m_b^2}{\left(y z
   \left(m_c^2-m_b^2\right)+m_b^2\right){}^2}+\frac{1}{y z
   \left(m_c^2-m_b^2\right)+m_b^2}\right),\\
s_{6}(m_b,m_c)&=\int^1_0dx\int^1_0dy\int^1_0dz6 y z^2 (2 x y z+2 y z-2 z-1)  \nonumber\\
&\cdot\left(\frac{m_b^2}{\left(y z
   \left(m_c^2-m_b^2\right)+m_b^2\right){}^2}+\frac{1}{y z
   \left(m_c^2-m_b^2\right)+m_b^2}\right),\\
s_{7}(m_b,m_c)&=\int^1_0dx\int^1_0dy\int^1_0dz12 y z^2 (y z-z)  \nonumber\\
&\cdot\left(\frac{m_b^2}{\left(y z
   \left(m_c^2-m_b^2\right)+m_b^2\right){}^2}
   +\frac{1}{y z \left(m_c^2-m_b^2\right)+m_b^2}\right),\\
s_{8}(m_a,m_b,m_c)&=\int^1_0dx\int^1_0dy\int^1_0dz\frac{6 y z^2 m_a m_b}{\left(x y z \left(m_a^2-m_c^2\right)+y z
   \left(m_c^2-m_b^2\right)+m_b^2\right){}^2},\\
s_{9}(m_a,m_b,m_c)&=\int^1_0dx\int^1_0dy\int^1_0dz\frac{6 y z^3 m_a m_b}{\left(x y z \left(m_a^2-m_c^2\right)+y z
   \left(m_c^2-m_b^2\right)+m_b^2\right){}^2},\\
s_{10}(m_a,m_b,m_c)&=\int^1_0dx\int^1_0dy\int^1_0dz\frac{6 y z^2 m_a m_b (-2 x y z+2 y z+1)}{\left(x y z \left(m_a^2-m_c^2\right)+y z
   \left(m_c^2-m_b^2\right)+m_b^2\right){}^2},\\
s_{11}(m_a,m_b,m_c)&=\int^1_0dx\int^1_0dy\int^1_0dz\frac{6 y z^2 m_a m_b (2 y z-2 z+1)}{\left(x y z \left(m_a^2-m_c^2\right)+y z
   \left(m_c^2-m_b^2\right)+m_b^2\right){}^2},\\
s_{12}(m_a,m_b,m_c)&=\int^1_0dx\int^1_0dy\int^1_0dz\frac{6 y z^2 m_a m_b}{\left(y z \left(m_a^2-m_b^2\right)+x y z
   \left(m_c^2-m_a^2\right)+m_b^2\right){}^2},\\
s_{13}(m_a,m_b,m_c)&=\int^1_0dx\int^1_0dy\int^1_0dz\frac{6 y z^3 m_a m_b}{\left(y z \left(m_a^2-m_b^2\right)+x y z
   \left(m_c^2-m_a^2\right)+m_b^2\right){}^2},\\
s_{14}(m_a,m_b,m_c)&=\int^1_0dx\int^1_0dy\int^1_0dz\frac{6 y z^2 m_a m_b (-2 x y z+2 y z+1)}{\left(y z \left(m_a^2-m_b^2\right)+x y z
   \left(m_c^2-m_a^2\right)+m_b^2\right){}^2},\\
s_{15}(m_a,m_b,m_c)&=\int^1_0dx\int^1_0dy\int^1_0dz\frac{6 y z^2 m_a m_b (2 y z-2 z+1)}{\left(y z \left(m_a^2-m_b^2\right)+x y z
   \left(m_c^2-m_a^2\right)+m_b^2\right){}^2}.
\end{align}


\end{document}